\documentclass[iop]{emulateapj}
\usepackage{apjfonts}
\usepackage{amssymb,latexsym,amsmath,graphics}
\slugcomment{} \shorttitle{Parallel tracks as quasi-steady states in neutron-star LMXBs}
\shortauthors{Erkut and \c{C}atmabacak}

\begin{document}

\title{Parallel tracks as quasi-steady states for the magnetic boundary layers in neutron-star low-mass X-ray binaries}

\author{M. Hakan Erkut\altaffilmark{1,2} and Onur \c{C}atmabacak\altaffilmark{3}}

\affil{\altaffilmark{1}Physics Engineering Department, Faculty of Science and
Letters, Istanbul Technical University, 34469, Istanbul, Turkey; mherkut@gmail.com}

\affil{\altaffilmark{2}Feza G\"{u}rsey Center for Physics and
Mathematics, Bo\u{g}azi\c{c}i University, 34684, \c{C}engelk\"{o}y,
Istanbul, Turkey}

\affil{\altaffilmark{3}Institute for Computational Sciences Y11 F74,
University of Zurich, Winterthurerstrasse 190, CH-8057 Zurich, Switzerland}

\begin{abstract}
The neutron stars in low-mass X-ray binaries (LMXBs) are usually thought to be weakly magnetized objects accreting matter from their low-mass companions in the form of a disk. Albeit weak as compared to those in young neutron-star systems, the neutron-star magnetospheres in LMXBs can play an important role in determining the correlations between spectral and temporal properties. Parallel tracks appearing in the plane of kilohertz (kHz) quasi-periodic oscillation (QPO) frequency versus X-ray flux can be used as a tool to study the magnetosphere-disk interaction in neutron-star LMXBs. For dynamically important weak fields, the formation of a non-Keplerian magnetic boundary layer at the innermost disk truncated near the surface of the neutron star is highly likely. Such a boundary region may harbor oscillatory modes of frequencies in the kHz range. We generate parallel tracks using the boundary region model of kHz QPOs. We also present the direct application of our model to the reproduction of the observed parallel tracks of individual sources such as 4U~1608--52, 4U~1636--53, and Aql~X-1.  We reveal how the radial width of the boundary layer must vary in the long-term flux evolution of each source to regenerate the parallel tracks. The run of the radial width looks similar for different sources and can be fitted by a generic model function describing the average steady behavior of the boundary region in the long term. The parallel tracks then correspond to the possible quasi-steady states the source can occupy around the average trend.
\end{abstract}


\section{Introduction}\label{intr}

The X-ray power spectra of low-mass X-ray binaries (LMXBs), where the accreting compact object is a neutron star, may exhibit millisecond quasi-periodic oscillations in the form of two simultaneous peaks usually being detected in the $200-1200$~Hz range \citep{Klis2000}. The peak separation between the centroid frequencies of these two kilohertz quasi-periodic oscillations (kHz QPOs) slightly varies around $300$~Hz throughout the neutron-star LMXB sources largely differing from each other in X-ray luminosity, spectral type, and neutron-star spin (or burst) frequency observed in the $200-600$~Hz range \citep{MB2007}.

The spectral classification of neutron-star LMXBs leads to the emergence of two main types of sources according to the criterion based on the identification of the source from the particular shape of its track in X-ray color-color and hardness-intensity diagrams \citep{HK1989}. These seemingly different types of sources, once compared to each other, appear as the brighter and dimmer LMXBs respectively, in X-ray luminosities, $L_{\mathrm{X}}$. For the so-called Z sources, $L_{\mathrm{X}}$ is close to the Eddington limit, $L_{\mathrm{E}}$. The atoll sources, on the other hand, are usually observed with $L_{\mathrm{X}}$ in the $0.005-0.2$~$L_{\mathrm{E}}$ range \citep{Ford2000}. The Z and atoll sources, however, cannot merely be distinguished by the shapes they trace out in X-ray color-color and hardness intensity diagrams. As in the case of 4U~1705--44, Z-shaped tracks can also be exhibited by the atoll sources. Moreover, a transient neutron-star LMXB, such as XTE~J1701--462, can evolve from Z source to atoll source behavior \citep{Homan2010,Sanna2010}.

Correlations between timing and spectral behaviors of both Z and atoll
sources are mainly determined by the track sources trace in X-ray
color-color diagrams. The curve-length parameters, $S_{\mathrm{Z}}$ and $S_{\mathrm{a}}$, are used to designate the particular location of the Z and atoll
sources respectively, along the track through different branches in the
color-color diagram \citep{Wijnands1997,Mendez1999}. The sense of increasing
curve-length parameters is usually interpreted as the sense of increasing inferred
mass accretion rate (inferred $\dot{M}$). In all Z-sources, the kHz QPOs have been
observed at the lowest inferred $\dot{M}$ levels. In atoll sources, these
high-frequency QPOs are usually detected for sufficiently high values of
inferred $\dot{M}$, which corresponds to the lower banana branch of the
U-shaped track in the color-color diagram \citep{Klis2000}.

The so-called parallel tracks have been observed in the plot of kHz QPO frequency, $\nu_{\mathrm{kHz}}$, versus X-ray count rate for individual sources \citep{Zhang1998,MK1999,Mendez1999,Mendez2000,Reig2000}. Each track in the plot describes the correlation between X-ray flux, $F_{\mathrm{X}}$, and $\nu_{\mathrm{kHz}}$ quite successfully on short time scales such as hours or less than a day. On longer time scales (more than a day), this correlation breaks down and a set of different parallel tracks appears in the plane of $\nu_{\mathrm{kHz}}$ versus $F_{\mathrm{X}}$ \citep{Zhang1998}.

In 4U~0614+09, the kHz QPOs were observed with different frequencies as compared to those observed 3 months later with similar count rates, although strong dependence of $\nu_{\mathrm{kHz}}$ on the count rate was found in each observation \citep{Ford1997}. The QPO frequency versus count rate diagram of the same source revealed the absence of a single correlation between these two quantities \citep{Mendez1997}. In nearly all observations of 4U~1608--52, \citet{Mendez1998} found that $\nu_{\mathrm{kHz}}$ is positively correlated to the count rate of the source, and yet the observations separated by several days within a month cannot be depicted by a simple function. The early analyses led to the conclusion that either the count rate is not a good measure of $\dot{M}$ or another quantity, at least beside $\dot{M}$, affects the QPO frequencies. The fact that $\nu_{\mathrm{kHz}}$ is not uniquely determined by $F_{\mathrm{X}}$ may also be a by-product of different regimes in the inner region of accretion disk. Each regime represented by a different track in the plane of $\nu_{\mathrm{kHz}}$ versus $F_{\mathrm{X}}$ may then correspond to a different vertical scale height during the evolution of the disk structure \citep{Zhang1998}.

In Z sources such as GX~340+0 and GX~5--1, twin simultaneous kHz QPOs were detected with frequencies increasing along the horizontal branch up to the upper part of the normal branch in the X-ray color-color diagram \citep{Jonker1998,Wijnands1998}. Using the parametrization to measure the position of the source along the Z track, $S_{\mathrm{Z}}$, unlike $F_{\mathrm{X}}$, was shown to correlate with $\nu_{\mathrm{kHz}}$ throughout the spectral evolution. In atoll sources, such as 4U~1608--52, 4U~1728--34, Aql~X-1, and 4U~0614+09, $\nu_{\mathrm{kHz}}$ correlates much better to $S_{\mathrm{a}}$, which is used to measure the position of the source along the atoll track \citep{Mendez1999,MK1999,Reig2000,Straaten2000}.

In an early plot of $\nu_{\mathrm{kHz}}$ versus X-ray count rate of the atoll source, 4U~1820--30, \citet{Zhang98b} reported the constancy of $\nu_{\mathrm{kHz}}$ above a certain value of the count rate. The dependence of $\nu_{\mathrm{kHz}}$ on both the energy flux and X-ray color, which are considered to be the possible indicators of $\dot{M}$ in 4U~1820--30, was also studied by \citet{Kaaret1999}. They found that $\nu_{\mathrm{kHz}}$ saturates at sufficiently high values of $\dot{M}$ in either indicators. This behavior was interpreted as the strong evidence of the inner disk being terminated by the marginally stable orbit. The saturation in the frequency versus flux relation of 4U~1820--30, however, may not be so apparent when a larger data set is used as there seems to be no evidence in other sources for a similar levelling off in frequency \citep{Klis2000}. In the lower kHz QPO frequency versus X-ray count rate diagram of Aql~X-1, the parallel tracks seem to have collapsed at high count rates, notwithstanding the source has attained even higher frequencies at much lower count rates \citep{Barret2008}.

The parallel tracks also appear in the plane of $\nu_{\mathrm{kHz}}$ versus $L_{\mathrm{X}}$ as far as the ensemble of LMXBs is concerned. The plane exhibits the frequency distribution of LMXB sources differing by orders of magnitude in $L_{\mathrm{X}}$ in the form of parallel-like groups. Almost the same frequency range, e.g., $500-1000$ Hz range, can be covered not only by the sources with $L_{\mathrm{X}}$ close to the Eddington luminosity, $L_{\mathrm{E}}$, but also by the sources with $L_{\mathrm{X}}\approx 10^{-2}L_{\mathrm{E}}$ \citep{Ford2000}.

The early attempts to explain the lack of correlation between $\nu_{\mathrm{kHz}}$ and $L_{\mathrm{X}}$ on long time scales in a given source came up with the proposal of slow variations in the accretion geometry, the relative contributions from different components such as disk and radial inflows, and the energetics of jets \citep{Klis1995,Wijnands1996,Kaaret1998,Mendez1999}. The parallel tracks, which are reminiscent of those observed in the frequency-count rate plot of the atoll source, 4U~1608--52, could be obtained within a scenario, where the instantaneous accretion rate through the disk,
$\dot{M}_{\mathrm{d}}$, and its long-term average determine both the QPO frequency and $L_{\mathrm{X}}$ \citep{Klis2001}. In analogy to the observed parallel tracks in the
$\nu_{\mathrm{kHz}}$ versus $L_{\mathrm{X}}$ diagram for the ensemble of sources, \citet{Klis2001} also simulated the frequency distributions of four sources differing in average
$\dot{M}_{\mathrm{d}}$.

In the peculiar atoll source, 4U~1636--53, significant shifts of the track the source follows in the color-color and hardness-intensity diagrams, similar to those occurring in some Z sources, were observed \citep{Salvo2003}. Unlike in other atoll sources, the parallel tracks in the
$\nu_{\mathrm{kHz}}$ versus intensity plot do not overlap perfectly in 4U~1636--53 when
$\nu_{\mathrm{kHz}}$ is plotted versus $S_{\mathrm{a}}$. The observed shifts in the atoll track were proposed by \citet{Salvo2003} as a possible reason behind the lack of strong correlation between the QPO frequencies and the color-color position of the source in the long term. To quantify the incidence of kHz QPOs in 4U~1636--53, \citet{MS2004} used the five years data of the source to obtain the histogram of the number of detections of QPOs as a function of count rate. The incidence of kHz QPOs was found to decrease as the flux level increases in the long term. While the parallel tracks coincide with the low flux levels, the number of QPOs drops sharply at intermediate flux levels. Similar behavior was also encountered in the $\nu_{\mathrm{kHz}}$ versus count rate diagrams for other sources, such as 4U~1608--52 and Aql~X-1, where the high-intensity tracks usually have a much narrower range of $\nu_{\mathrm{kHz}}$ as compared to the low-intensity tracks \citep{Mendez2000,Barret2008}.

Beside the spectral shape, the long-term intensity may also be crucial in understanding the properties of kHz QPOs \citep{MS2004}. The possible existence of an inferred threshold $\dot{M}$, beyond which QPOs seem to be suppressed, can be revealed by the long-term intensity evolution of the source. If there is a threshold $\dot{M}$ in the long-term flux evolution of the source as suggested by the work of \citet{MS2004}, the flux can then be a good indicator of $\dot{M}$ not only in the short term but also in the long term. The lack of correlation between $\nu_{\mathrm{kHz}}$ and $F_{\mathrm{X}}$ in the long term can be due to another parameter or parameters determining $\nu_{\mathrm{kHz}}$ in addition to $\dot{M}$. The long-term average of $\dot{M}_{\mathrm{d}}$ was proposed by \citet{Klis2001} as a second parameter that determines both $\nu_{\mathrm{kHz}}$ and $L_{\mathrm{X}}$ in addition to $\dot{M}_{\mathrm{d}}$. In this model, the observed variation in $F_{\mathrm{X}}$ is due to changes in both the instantaneous mass inflow rate in the accretion disk and its long-term average. The model can yield the disconnected tracks if the observational windowing is simulated by gaps between numerical data sets. If, however, the distinct tracks correspond to certain flux levels, at which the kHz QPOs can be produced, the gaps may not only arise from undersampling \citep{MS2004}. The direct observation of several transitions between the two parallel tracks of 4U~1636--53 by \citet{Mendez2003} could be the first evidence that kHz QPOs can only be generated at some definite flux levels. The timescale of these fast transitions is smaller than the total time the source spends on a given track. Each track may thus represent a near-equilibrium or quasi-steady state, where the variations in $\dot{M}_{\mathrm{d}}$ are sufficiently slow in order for the parameter responsible for $\nu_{\mathrm{kHz}}$ to acquire definite values. The transitions between different tracks, on the other hand, might be triggered due to fast variations in $\dot{M}_{\mathrm{d}}$. The source then seeks another track to settle down to a near-equilibrium state for the generation of kHz QPOs.

The distribution of lower kHz QPO frequencies with respect to accretion-related parameters has recently been studied for the ensemble of 15 LMXBs \citep{Erkut2016}. Based on the possibility that the magnetic field strength, $B$, on the surface of the neutron star differs from one source to another, \citet{Erkut2016} have shown that the existence of a correlation between the lower kHz QPO frequency, $\nu_{1}$, and $\dot{M}/B^2$, is very likely. The lack of correlation between $\nu_{\mathrm{kHz}}$ and $L_{\mathrm{X}}$ in the same ensemble \citep{Ford2000}, can therefore be accounted for, at least to some extent, by the new correlation between $\nu_{1}$ and $\dot{M}/B^2$. The frequency distribution in the $\nu_{1}$ versus $\dot{M}/B^2$ plane for the ensemble of sources can be successfully described by the cumulative effect of the model function fits to data of individual sources.

The model function for $\nu_{1}$ can account for the data tracks of different slopes in a given source within the ensemble, provided that the radial width of the inner boundary region of the accretion disk changes as $\dot{M}$ varies in time. According to \citet{Erkut2016}, the variation of the width of the boundary region with $\dot{M}$ can be estimated by a physically motivated function, i.e., the aspect ratio of the disk. In the absence of sufficiently high number of data points per source, however, it is difficult to distinguish the fit quality of the physically motivated function from that of a simple linear or power-law function \citep{Erkut2016}. In this paper, we address the observed phenomenon of parallel tracks in individual neutron-star LMXB sources within the boundary region model of kHz QPOs \citep{AP2008,EPA2008}. We focus on the parallel tracks with sufficiently high number of data points in the $\nu_{1}$ versus X-ray count rate diagram of individual sources such as 4U~1608--52, 4U~1636--53, and Aql~X-1.

The next section (Section~\ref{theory}) is devoted to a brief summary of the theory of interaction between the neutron-star magnetosphere and the innermost disk matter. In this section, the model function for $\nu_{1}$ is derived from first principles and arguments that are similar to those in \citet{Erkut2016}. The present function, however, contains more details regarding the physics of interaction throughout the boundary region in the inner disk. In Section~\ref{anlys}, we exploit the model function to mimic the parallel tracks by studying the variations of the boundary region width and other physical parameters with the mass inflow rate in the accretion disk. We then reproduce the observed tracks numerically in the plane of $\nu_{1}$ versus X-ray flux for individual sources and obtain the evolution of the model parameters with $\dot{M}$. The outcomes of our analysis in Section~\ref{anlys} are interpreted and discussed in Section~\ref{disc}.

\section{Theory}\label{theory}

The interaction between the magnetosphere of a star and its accretion disk has been the subject of an ongoing debate for more than 40 years. Although it is still under investigation as far as the extreme regimes, such as the propeller and super-Eddington phases, are concerned, the radius, where the inner disk is truncated by the neutron-star magnetosphere, is fairly established by convention and can be estimated using the balance between magnetic and ram pressures or magnetic and material stresses \citep{GL1979}.

Throughout the present study, we work with cylindrical coordinates. Using the typical values (denoted by a subscript $t$) of the magnetic field, the density of matter, and the radial drift and azimuthal velocities at the magnetopause, we write
\begin{equation}
\frac{B_t^2}{8\pi} = \frac{1}{2} \rho_t v_{r,t} v_{\phi,t}, \label{balance}
\end{equation}
to estimate the typical value of the plasma beta,
\begin{equation}
\beta_t =\frac{4\pi \rho_t c^2_{s,t}}{B^2_t}, \label{beta}
\end{equation}
in the magnetically dominated boundary region situated at the innermost disk radius,
\begin{equation}
r_{\mathrm{in}} \simeq r_{\mathrm{A}}=\dot{M}^{-2/7} B^{4/7} \left( GM \right)^{-1/7} R^{12/7}, \label{rin}
\end{equation}
where $r_{\mathrm{A}}$ is the Alfv\'{e}n radius, which we express in terms of the mass, $M$, radius, $R$, and surface magnetic dipole field strength, $B$, of the neutron star. The typical value of the sound speed in the disk can be estimated as
\begin{equation}
c_{s,t}=\Omega_{\mathrm{K}}\left( r_{\mathrm{in}} \right)H_t=\varepsilon v_{\phi,t}, \label{sound}
\end{equation}
using the typical half-thickness, $H_t$, of the disk and the typical value of the azimuthal velocity, which we define in terms of the Keplerian orbital velocity at the innermost disk radius, that is,
\begin{equation}
v_{\phi,t}=\Omega_{\mathrm{K}}\left( r_{\mathrm{in}} \right)r_{\mathrm{in}}. \label{azimut}
\end{equation}
As seen in Equation~(\ref{sound}), the typical speed of orbital motion is as supersonic as it is in a standard thin disk \citep{SS73}, i.e., $c_{s,t}<v_{\phi,t}$ by a factor of the typical aspect ratio of the disk,
\begin{equation}
\varepsilon=\frac{H_t}{r_{\mathrm{in}}}, \label{epsilon}
\end{equation}
which we introduce as a small parameter into the magnetohydrodynamic equations in Section~\ref{rdbr} to reveal the existence of a magnetic boundary region at $r_{\mathrm{in}}$.
The efficient angular momentum transport within a magnetic boundary region is due to magnetic stresses exerted on the disk matter by the large scale magnetic fields of stellar origin. As the angular momentum of the inner disk matter is removed by the magnetosphere, the so-called magnetic braking is characterized by a sharp decrease in the density of the inner disk, where the radial drift velocity exceeds its typical value in the outer disk and becomes comparable in magnitude with the speed of sound \citep{GL1979,RUKL02,EA2004}. Choosing therefore $v_{r,t}=c_{s,t}$ and making use of Equations~(\ref{balance}), (\ref{beta}), and (\ref{sound}), we find $\beta_t=\varepsilon$ for the plasma beta in the magnetic boundary region.

\subsection{Rotational Dynamics of the Boundary Region}\label{rdbr}

We assume that the mass accretion rate onto the neutron star is directly determined by the mass inflow rate in the accretion disk. We therefore estimate the typical mass accretion rate as
\begin{equation}
\dot{M}=\dot{M}_{\mathrm{d}}=4\pi H_t r_{\mathrm{in}}\rho_t c_{s,t}, \label{mdot}
\end{equation}
using $v_{r,t}=c_{s,t}$ and the constant mass influx condition following the continuity equation. The angular momentum balance for a steady magnetic boundary region can be written as
\begin{equation}
\frac{d}{dr}\left(\dot{M}r^2\Omega \right)=-r^2 B^+_{\phi}B_z, \label{angmom}
\end{equation}
neglecting the viscous stresses in the $\phi$-component of the momentum conservation \citep{GL1979,EA2004}. Here, $B^+_{\phi}\equiv B_{\phi}(z=H)$ and $B_z$ are the surface toroidal and dominant poloidal components of the large scale magnetic field threading the boundary region, respectively. The angular velocity of the matter, $\Omega$, is expected to deviate from its Keplerian value due to the sufficiently large magnetic pressure gradients in the same region. To estimate the radial profile of $B_z$, and thus the magnetic pressure gradients in the boundary region, we consider, for $\left|v_z\right|\ll \left|v_r\right|$, the poloidal component of the induction equation \citep{EA2004},
\begin{equation}
\frac{\partial B_r}{\partial z}-\frac{\partial B_z}{\partial r}=-\frac{v_r B_z}{\eta}, \label{ind}
\end{equation}
where $\eta$ is the effective magnetic diffusivity in the boundary region. Choosing $\eta=-D_{\mathrm{BL}}v_r\lambda r_{\mathrm{in}}$ with $D_{\mathrm{BL}}$ being the diffusivity coefficient of order unity, it follows from the vertical integration of Equation~(\ref{ind}) that
\begin{equation}
\frac{dB_z}{dr}=-\left(1-\frac{D_{\mathrm{BL}}\lambda r_{\mathrm{in}}}{H}\frac{B^+_r}{B_z}\right)\frac{B_z}{D_{\mathrm{BL}}\lambda r_{\mathrm{in}}}, \label{vind}
\end{equation}
where the radial extension of the boundary region, $\delta r_{\mathrm{in}}\approx D_{\mathrm{BL}}\lambda r_{\mathrm{in}}$, which acts as the electromagnetic screening length for the poloidal magnetic field, and the vertical disk scale height, $H$, are expected to be of the same order of magnitude \citep{GL1979}. For $\left|B^+_r\right|\ll\left|B_z\right|$, Equation~(\ref{vind}) can be reduced to
\begin{equation}
\frac{dB_z}{dr}=-\frac{B_z}{D_{\mathrm{BL}}\lambda r_{\mathrm{in}}} \label{svind}
\end{equation}
whose solution corresponds to the efficient screening of the poloidal field by the toroidal electric currents flowing across the boundary region.

Next, we normalize each quantity by its typical value and express Equations~(\ref{angmom}) and (\ref{svind}) in dimensionless form. We define $x=r/r_{\mathrm{in}}$, $\omega=\Omega/\Omega_{\mathrm{K}}\left( r_{\mathrm{in}} \right)$, $b_z=B_z/B_t$, and the azimuthal pitch,
$\gamma_{\phi}=B^+_{\phi}/B_z<0$, in the boundary region to obtain
\begin{equation}
\beta_t\frac{d}{dx}\left(x^2\omega \right)=\left|\gamma_{\phi}\right|x^2 b^2_z \label{ndam}
\end{equation}
and
\begin{equation}
\frac{db_z}{dx}=-\frac{b_z}{D_{\mathrm{BL}}\lambda} \label{ndin}
\end{equation}
with the help of Equations~(\ref{beta}), (\ref{sound}), and (\ref{mdot}). Here, the dimensionless length scale, $\lambda$, plays the role of the so-called distinguished limit of the boundary layer problem. To solve Equations~(\ref{ndam}) and (\ref{ndin}) in the neighborhood of $r=r_{\mathrm{in}}$ ($x=1$), we perform the coordinate stretching, $x=1+\lambda X$, where $X$ is a new variable to measure how fast the physical quantities change along the radial direction in the boundary region \citep{Regev1983}. Using $\beta_t=\varepsilon$, we rewrite Equations~(\ref{ndam}) and (\ref{ndin}) for $\lambda \ll 1$ as
\begin{equation}
\frac{\varepsilon}{\lambda}\frac{d\omega}{dX}=\left|\gamma_{\phi}\right|b^2_z \label{blam}
\end{equation}
and
\begin{equation}
\frac{db_z}{dX}=-\frac{b_z}{D_{\mathrm{BL}}}, \label{blin}
\end{equation}
respectively, in terms of the new variable. The solution of Equation~(\ref{blin}) can be found as
\begin{equation}
b_z\left(X\right)=b_0 e^{-X/D_{\mathrm{BL}}}, \label{slnbz}
\end{equation}
where $\left|b_0\right|$ is of order unity. The distinguished limit, on the other hand, can easily be guessed from Equation~(\ref{blam}) as $\lambda=\varepsilon$. Substituting Equation~(\ref{slnbz}) for $b_z$ into Equation~(\ref{blam}), we find
\begin{equation}
\omega_{\mathrm{BL}}\left(X\right)=1-\left(1-\omega_0\right)e^{-2X/D_\mathrm{BL}} \label{omega}
\end{equation}
for the angular velocity of the matter in the boundary layer. The innermost boundary condition for $\omega$ is given by
\begin{equation}
\omega_0\equiv \omega_{\mathrm{BL}}\left(X=0\right)=1-\frac{1}{2}\left|\gamma_{\phi}\right|D_{\mathrm{BL}}b^2_0. \label{omgzr}
\end{equation}
The dimensionless angular velocity of the Keplerian disk outside the boundary layer is
\begin{equation}
\omega_{\mathrm{d}} \left(x\right)=\frac{\Omega_{\mathrm{K}}(r)}{\Omega_{\mathrm{K}}(r_{\mathrm{in}})}=\left(\frac{r_{\mathrm{in}}}{r}\right)^{3/2}=x^{-3/2}. \label{omgk}
\end{equation}
According to the method of matched asymptotic expansions \citep{Regev1983}, the unified solution for the angular velocity can be obtained as
\begin{equation}
\omega \left(x\right)=\omega_{\mathrm{d}} \left(x\right)+\omega_{\mathrm{BL}}\left(X\right)-\omega_{\mathrm{d}} \left(x\rightarrow 1 \right), \label{unif}
\end{equation}
where the inner (boundary layer) and outer (disk) solutions match asymptotically, that is,
\begin{equation}
\omega_{\mathrm{BL}}\left(X\rightarrow \infty\right)=1=\omega_{\mathrm{d}} \left(x\rightarrow 1 \right), \label{match}
\end{equation}
as expected. Using Equations~(\ref{omega}), (\ref{omgk}), and (\ref{match}), it follows from Equation~(\ref{unif}) that
\begin{equation}
\omega \left(x\right)=x^{-3/2}-\left(1-\omega_0\right)\exp \left(-\frac{x-1}{\delta}\right). \label{bromg}
\end{equation}
Here, $\delta=\varepsilon D_{\mathrm{BL}}/2$, as a dimensionless parameter, represents the radial width of the boundary region. We note that the unified solution in Equation~(\ref{bromg}) includes the basic parameters such as the coefficient of magnetic diffusivity, the azimuthal pitch, and the aspect ratio of the disk at the magnetospheric radius and hence describes, in terms of magnetosphere-disk interaction, the rotational dynamics of the boundary region.

\subsection{Model Function for kHz QPO Frequencies}\label{qpomf}

According to the boundary region model, kHz QPOs from neutron stars are
interpreted as global hydrodynamic modes in the boundary layers of
accretion disks \citep{EPA2008}. The highest dynamical frequency in the
inner disk was first noted by \citet{AP2008} to be the radial epicyclic frequency $\kappa$. In a
sub-Keplerian boundary region, $\kappa$ exceeds both the local Keplerian angular
frequency, $\Omega _{\mathrm{K}}$, and the orbital angular frequency, $\Omega$. The stability analysis of the global modes in the inner disk has revealed that the modes with frequency bands around $\kappa$ and $\kappa \pm \Omega$ grow in amplitude as a result of the instability driven by surface density gradients in a sub-Keplerian boundary
region \citep{EPA2008}. The extension of the model to hectoHz QPOs in black hole systems is also possible provided general relativistic effects on the inner disk are taken into account \citep{Erkut2011}.

\begin{figure}
\epsscale{1.16} \plotone{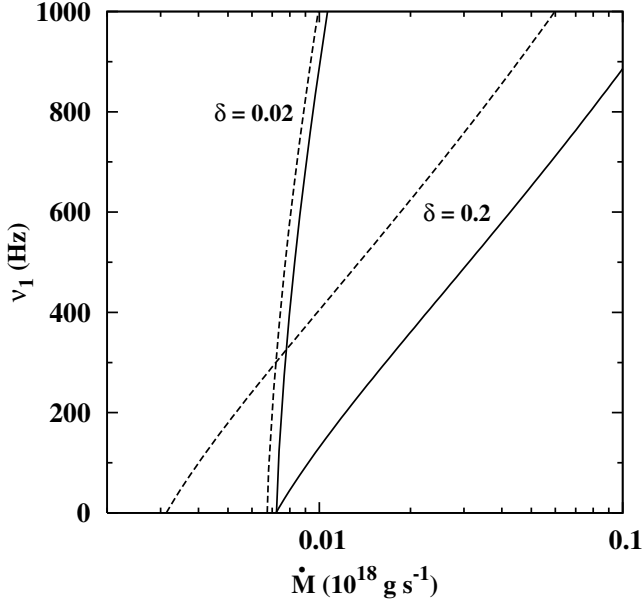} \caption{Lower kHz QPO frequency estimated by the model function in Equation~(\ref{nu1}) (solid curves) and the model function in \citet{Erkut2016} (dashed curves), assuming $\Delta \nu=300$~Hz for a neutron star of mass $M=1.4M_\odot$, radius $R=10$~km, and surface magnetic dipole field strength $B=10^8$~G. The solid and dashed curves with $\delta=0.02$ are steeper than those with $\delta=0.2$. The two model functions can be seen to be degenerate for sufficiently small values of $\delta$. For sufficiently large values of $\delta$, on the other hand, the model function in Equation~(\ref{nu1}) (solid curve) estimates lower values for kHz QPO frequency at a given $\dot{M}$ as compared to the model function in \citet{Erkut2016} (dashed curve). \label{fig1}}
\end{figure}

Throughout the present work, we identify the lower and upper kHz QPO frequencies using
$2\pi \nu_{1}=\kappa-\Omega$ and $2\pi \nu_{2}=\kappa$, respectively. The frequency difference between the two kHz QPO peaks is $\Delta \nu=\Omega/2\pi$. We normalize the radial epicyclic frequency,
\begin{equation}
\kappa \left(r\right)=\sqrt{2\Omega \left(2\Omega+r\frac{d\Omega}{dr}\right)}, \label{repic}
\end{equation}
with respect to $\Omega_{\mathrm{K}}\left(r_{\mathrm{in}}\right)$ and substitute Equation~(\ref{bromg}) into Equation~(\ref{repic}) to obtain
\begin{equation}
\kappa \left(r_{\mathrm{in}}\right)=\Omega_{\mathrm{K}}\left(r_{\mathrm{in}}\right)\sqrt{2\omega _0\left(2\omega _0 -\frac{3}{2}+\frac{1-\omega _0}{\delta}\right)} \label{kaparin}
\end{equation}
at $x=1$. We calculate the model function for the lower kHz QPO frequency,
\begin{equation}
\left. \nu_1=\frac{\kappa-\Omega}{2\pi} \right|_{r=r_{\mathrm{in}}}, \label{mdfnc}
\end{equation}
using Equation~(\ref{kaparin}). Remembering that the innermost boundary condition is defined by
\begin{equation}
\omega_0\equiv \frac{\Omega \left(r_{\mathrm{in}}\right)}{\Omega_{\mathrm{K}}\left(r_{\mathrm{in}}\right)}=\frac{\Delta \nu}{\nu_{\mathrm{K}} \left(r_{\mathrm{in}}\right)}, \label{bc}
\end{equation}
it follows from Equation~(\ref{mdfnc}) that
\begin{equation}
\nu_1=f\left(\nu_{\mathrm{A}},\delta,\Delta \nu \right)\sqrt{\nu_{\mathrm{A}}\Delta \nu}, \label{nu1}
\end{equation}
where $f$ is a dimensionless function of $\delta$, $\Delta \nu$, and the Keplerian frequency at the Alfv\'{e}n radius, $\nu_{\mathrm{A}}\equiv \nu_{\mathrm{K}} \left(r_{\mathrm{A}}\right)=\nu_{\mathrm{K}} \left(r_{\mathrm{in}}\right)$. We write $f\left(\nu_{\mathrm{A}},\delta,\Delta \nu \right)$ explicitly as
\begin{equation}
f=\sqrt{\frac{2}{\delta}}\left[\sqrt{1-\frac{3\delta}{2}-\left(1-2\delta \right)\frac{\Delta \nu}{\nu_{\mathrm{A}}}}-\sqrt{\frac{\delta \Delta \nu}{2\nu_{\mathrm{A}}}}\right]. \label{df}
\end{equation}
Note that the dimensionless function $f\left(\nu_{\mathrm{A}},\delta,\Delta \nu \right)$ in Equation~(\ref{nu1}), which is defined by Equation~(\ref{df}), is almost the same as
$F\left(\nu_{\mathrm{A}},\delta,\Delta \nu \right)$ in \citet{Erkut2016}. The only difference between $f$ and $F$ is the term $3\delta /2$ in the definition of $f$. The presence of this extra term in Equation~(\ref{df}) arises from matching the non-Keplerian rotation profile of the boundary region with the Keplerian disk. In Figure~\ref{fig1}, we compare the behavior of the model function in the present work (Equation~(\ref{nu1})) for the lower kHz QPO frequency as a function of $\dot{M}$ with that of the model function in \citet{Erkut2016}. Note that both functions yield almost the same frequencies for sufficiently low values of $\delta$. As $\delta$ gets bigger, the model function we derive in this paper (solid curve in Figure~\ref{fig1}) estimates lower values for $\nu_1$ in comparison with the model function employed by \citet{Erkut2016} (dashed curve in Figure~\ref{fig1}). Apart from the effect of $\delta$-parameter, which leads to the parallel shift of one model curve with respect to the other, the behaviors of the two functions are the same in the sense that the slopes of both functions increase as $\delta$ decreases.

The existence of a possible correlation between $\nu_1$ and the accretion-related parameter, $\dot{M}/B^2$, in the ensemble of neutron-star LMXBs has been revealed via the use of the model function shown by the dashed curves in Figure~\ref{fig1} \citep{Erkut2016}. The model function fit to the frequency data of each source in the ensemble defines a region spanned by a set of model curves within a given range of $\delta$. The frequency distribution for the ensemble of sources in the $\nu_1$ versus $\dot{M}/B^2$ plane is then obtained through the merger of different regions of the model function fit to individual source data for similar neutron-star masses and radii. An example of such a region would be the one bounded by the model curves labeled with $\delta=0.02$ and $\delta=0.2$ in Figure~\ref{fig1}. For the maximum value of $\delta$ in a given range, e.g., $\delta=0.2$, the model function in Equation~(\ref{nu1}) (solid curve in Figure~\ref{fig1}) would define a wider region as compared to the model function in \citet{Erkut2016}. The observed distribution of kHz QPO frequencies in individual sources and therefore the same correlation with $\nu_1$ in the ensemble of sources could then be accounted for choosing smaller values for the maximum $\delta$ if the model function in Equation~(\ref{nu1}) were used instead of the model function in \citet{Erkut2016}.

In addition to its advantage of covering a wider region of fit to QPO data within the same range of $\delta$, the model function we introduce in Equation~(\ref{nu1}) has well defined parameters such as $\delta=\varepsilon D_{\mathrm{BL}}/2$ and $\Delta \nu/\nu_{\mathrm{A}}$, which are related to the aspect ratio of the disk and the azimuthal pitch in the boundary region through Equations~(\ref{epsilon}), (\ref{omgzr}), and (\ref{bc}). We devote the next section to the implications of the model function in Equation~(\ref{nu1}) for the observed properties of parallel tracks. First we explore the effects of variations of model parameters on the distribution and slopes of the tracks, then we use the model function to reproduce the parallel tracks of three LMXBs, namely, 4U~1608--52, 4U~1636--53, and Aql~X-1, for which the long-term evolution of the X-ray flux is available.

\section{Analysis}\label{anlys}

\subsection{Implications of Model Parameters for Parallel Tracks}\label{mifpt}

\begin{figure}
\epsscale{1.15} \plotone{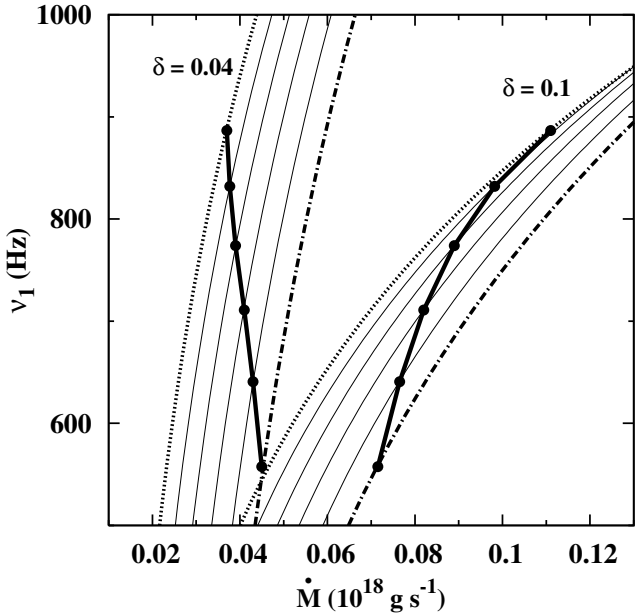} \caption{Lower kHz QPO frequency estimated by the model function in Equation~(\ref{nu1}) for two different values of $\delta$ when $\Delta \nu$ varies in the $220-320$~Hz range according to the observed correlation between the two kHz QPO frequencies in Sco~X-1 (Equation~(\ref{ffcor})). The plot is for a neutron star of mass $M=1.4M_\odot$, radius $R=10$~km, and surface magnetic dipole field strength $B=2\times10^8$~G. For both $\delta=0.04$ and $\delta=0.1$, the solution of Equation~(\ref{ffcor}) for $\nu_1$ is shown by a series of black dots, each of which lies on a model curve labeled with a different $\Delta \nu$ changing from $320$~Hz (dot dashed curve) to $220$~Hz (dotted curve). The black dots are connected to each other through thick solid lines to form a track across the model curves with constant $\delta$. All model curves labeled with intermediate values between $320$~Hz and $220$~Hz are depicted by thin solid curves. \label{fig2}}
\end{figure}

The peak separation $\Delta \nu=\nu_2-\nu_1$ between the two kHz QPO frequencies usually decreases when $\nu_1$ and $\nu_2$ increase \citep{Klis2000}. In Sco~X-1, the observed correlation between $\nu_1$ and $\nu_2$ \citep{Psaltis1998} can be written, to a good approximation, as
\begin{equation}
\frac{\nu_1}{724\, \mathrm{Hz}}\simeq \left(\frac{\nu_1+\Delta \nu}{1000\, \mathrm{Hz}}\right)^2. \label{ffcor}
\end{equation}
For illustrative purposes, we use Equation~(\ref{ffcor}) to compute $\nu_1$ as a function of $\Delta \nu$ and show, in Figure~\ref{fig2}, the six values of $\nu_1$ corresponding to the $220-320$~Hz range for $\Delta \nu$. The model function in Equation~(\ref{nu1}) is plotted as two sets of curve for $\delta=0.04$ and $\delta=0.1$. In each set, $\Delta \nu$ varies from 320~Hz (dot dashed curve) to 220~Hz (dotted curve) with intermediate values being drawn by thin solid curves. The solution of Equation~(\ref{ffcor}) for $\nu_1$ is represented by six black dots connected to each other through thick solid lines crossing the model function curves of a given set with constant $\delta$. We obtain Figure~\ref{fig2} for a putative neutron star of mass $M=1.4M_\odot$, radius $R=10$~km, and surface magnetic dipole field strength $B=2\times10^8$~G. By adopting smaller $M$ and larger $R$ and $B$ values, we could obtain a plot similar to Figure~\ref{fig2}, but for relatively larger values of $\dot{M}$. Note that changing $\Delta \nu$, while keeping $\delta$ constant may result in tracks with negative slopes in the $\nu_1$ versus $\dot{M}$ plane for sufficiently small values of $\delta$ and vice versa. As the slope of the model curve with small enough $\delta$ is higher than that with a larger $\delta$, the source can occupy the same range of $\nu_1$ even if the change in $\dot{M}$ or count rate is limited to relatively shorter timescales as compared to larger variations in $\dot{M}$ when $\delta$ is large (Figure~\ref{fig2}). Moreover, the slope of the track in the plane of $\nu_{\mathrm{kHz}}$ versus X-ray flux can be negative as in the case of Sco~X-1, where $\Delta \nu$ decreases as both kHz QPO frequencies increase, if $\delta$ is small enough as shown in Figure~\ref{fig2} \citep{Yu2001,Boutelier2010}.

\begin{figure}
\epsscale{1.16} \plotone{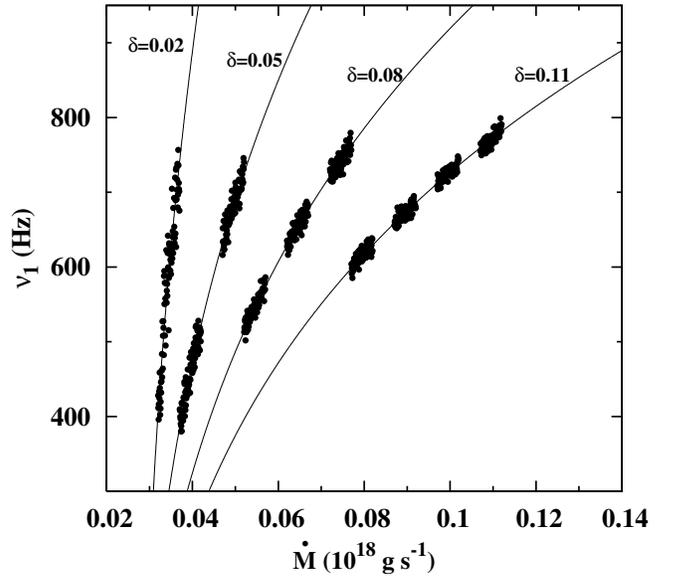} \caption{Lower kHz QPO frequency estimated by the model function in Equation~(\ref{nu1}) as $\delta$ acquires values changing from 0.02 to 0.11 while $\Delta \nu=300$~Hz remains constant. The plot is for a neutron star of mass $M=1.4M_\odot$, radius $R=10$~km, and surface magnetic dipole field strength $B=2\times10^8$~G. Each solid curve labels the model function with a reference value for $\delta$. The tracks (dark spots) consist of numerical data generated on each curve when $\delta$ is allowed to fluctuate randomly about its reference value. Disconnected tracks along a curve labeled with a particular $\delta$ are obtained through sampling of numerical data sets, which are separated from each other by $\dot{M}$-intervals of width $5\times 10^{15}\,\mathrm{g\,s}^{-1}$ to mimic the effect of observational windowing. \label{fig3}}
\end{figure}

Unlike the Z-sources such as Sco~X-1, where the frequency range, over which $\nu_1$ and $\nu_2$ are detected simultaneously, is relatively narrow $(\lesssim 300\, \mathrm{Hz})$, the atoll sources such as 4U~1608--52 and 4U~1636--53 exhibit kHz QPOs in a wider range $(\gtrsim 400\, \mathrm{Hz})$ of frequencies throughout the long-term flux evolution \citep{Boutelier2010}. The distribution of kHz QPO frequencies includes parallel tracks of different slopes, which can be accounted for by the ever-increasing values of $\delta$ as $L_{\mathrm{X}}$ or $\dot{M}$ increases \citep{Erkut2016}. To see the effect of variation in $\delta$ on the distribution and slopes of parallel tracks, we plot $\nu_1$ as a function of $\dot{M}$ for different values of $\delta$, however, keeping $\Delta \nu$ constant at 300~Hz as shown in Figure~\ref{fig3}. The slope of the model function (solid curves) can be seen to be positive for a given $\delta$ and larger for a smaller $\delta$ at a given $\dot{M}$. We let each $\delta$ fluctuate randomly about its reference value and generate the tracks in Figure~\ref{fig3} through selection of numerical data sets for $\nu_1$ to account for the effect of observational windowing along the same curve. The tracks appear as parallel spots because they are distributed over different values of $\delta$ ranging from 0.02 to 0.11. It is obvious from Figures~\ref{fig2} and \ref{fig3} that parallel tracks can only be obtained unless we keep both $\Delta \nu$ and $\delta$ constant. Changing, however, $\Delta \nu$ while keeping $\delta$ constant is not enough to produce the parallel tracks of atoll sources, which differ considerably from each other in both slope and count rate during the long-term flux evolution unlike the Z source Sco~X-1.

As we shall see later in this section, the interpretation of the observed parallel tracks in the atoll sources, 4U~1608--52, 4U~1636--53, and Aql~X-1, within the boundary region model of kHz QPOs requires the model parameter $\delta$ to vary in accordance with the theory of disk accretion. As $\dot{M}$ increases, $\delta$ is also expected to increase, at least for a certain range of $\dot{M}$, simply because it is proportional (see Section~\ref{theory}) to the typical aspect ratio of the disk, $\varepsilon=H_t/r_{\mathrm{in}}$, in the boundary layer \citep{Erkut2016}. In Figure~\ref{fig3}, we illustrate, without specifying how $\delta$ should change in the long-term $\dot{M}$ or X-ray flux evolution of the source, the distribution of the tracks on the model curves from $\delta=0.02$ to $\delta=0.11$ as $\dot{M}$ increases. We assume a certain range of $\dot{M}$, for which the system can pursue a particular $\delta$. Once the specified range of $\dot{M}$ is surpassed, the source jumps from one $\delta$-value to another. Note that our restriction for generating tracks in the high $\dot{M}$ regime with relatively high values of $\delta$ results in an effective suppression of QPOs as if there is a threshold $\dot{M}$ at sufficiently high flux levels. Such a behavior has been observed before in the atoll sources 4U~1608--52, 4U~1636--53, and Aql~X-1, where the $\nu_{\mathrm{kHz}}$ range of high-intensity tracks is narrower than that of low-intensity tracks \citep{Mendez2000,MS2004,Barret2008}.

\begin{figure}
\epsscale{1.16} \plotone{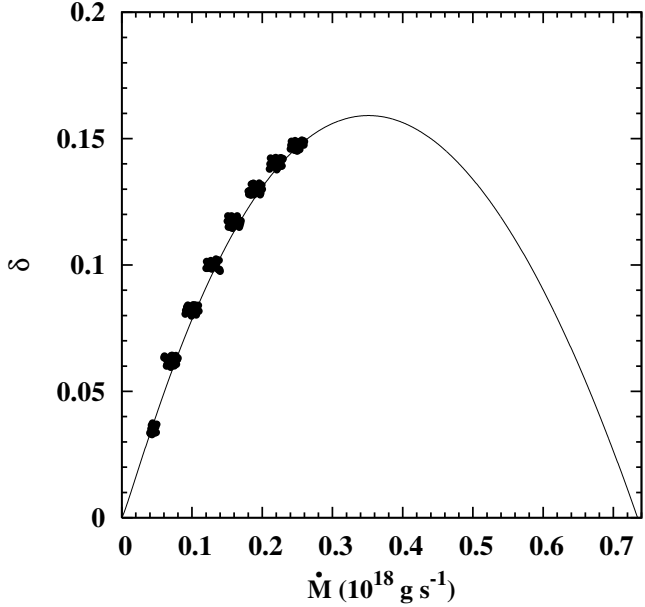} \caption{Dimensionless radial width of the boundary region, $\delta$, as a function of $\dot{M}$ for a neutron star of mass $M=2M_\odot$, radius $R=10$~km, and surface magnetic dipole field strength $B=3\times10^8$~G. Solid curve represents the behavior of $\delta$ when the disk is in steady state. The steady solution for $\delta$ is modeled using $H_t=aH_{\mathrm{SS}}\left(r_{\mathrm{A}}\right)+H_0$ in $\delta=\varepsilon D_{\mathrm{BL}}/2$, where $\varepsilon=H_t/r_{\mathrm{A}}$. The plot is obtained for $D_{\mathrm{BL}}=5$, $a=2.5$, $H_0=0$, and $C=1$. Here, $C$ is the angular momentum efficiency constant of order unity in $H_{\mathrm{SS}}\left(r_{\mathrm{A}}\right)$, i.e., the half-thickness of the Shakura-Sunyaev disk at $r_{\mathrm{A}}$ \citep{Erkut2016}. The dark spots designate the random fluctuation of $\delta$ about the steady solution estimated by the solid curve at the mean value of the corresponding $\dot{M}$-interval. \label{fig4}}
\end{figure}

\begin{figure}
\epsscale{1.16} \plotone{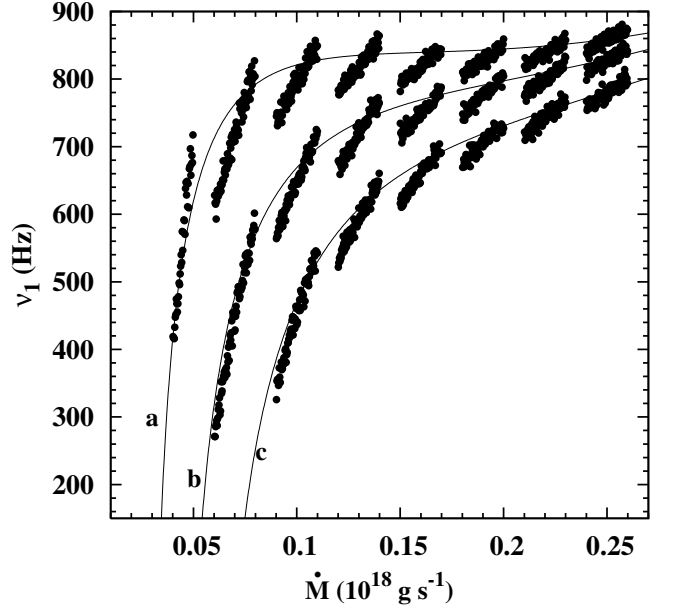} \caption{Lower kHz QPO frequency estimated by the model function in Equation~(\ref{nu1}) for a neutron star of $M=2M_\odot$, $R=10$~km, and $B=3\times10^8$~G for which $\delta$ varies with $\dot{M}$ as in Figure~\ref{fig4}. Thin solid curves labeled with a, b, and c would describe the behavior of the model function for $\Delta \nu=290$, 350, and 400~Hz, respectively if $\delta$ followed its steady-state trajectory (solid curve in Figure~\ref{fig4}). The parallel tracks on each curve correspond to the evaluation of the model function at $\delta$-values associated with the dark spots in Figure~\ref{fig4}, where $\delta$ fluctuates randomly about its steady-state value. \label{fig5}}
\end{figure}

We now specify how $\delta$ can vary as a function of $\dot{M}$ in a steady-state accretion disk. As we define at the end of Section~\ref{rdbr}, $\delta=\varepsilon D_{\mathrm{BL}}/2$, where $\varepsilon=H_t/r_{\mathrm{A}}$ is known as the typical aspect ratio of the disk (Equation~(\ref{epsilon})). In the absence of a model for the variation of magnetic diffusivity with $\dot{M}$, we keep $D_{\mathrm{BL}}$ constant to estimate the long-term evolution of $\delta$ with $\dot{M}$. As we shall discuss in Section~\ref{opt}, however, the value of $D_{\mathrm{BL}}$, which we deduce from observational data, may change from one parallel track to another, yet fluctuating around a constant mean value. We adopt the simple and plausible assumption in \citet{Erkut2016} that the typical half-thickness of the disk is in correlation with the vertical scale-height of the standard (Shakura-Sunyaev) disk at $r_{\mathrm{A}}$, i.e., $H_t=aH_{\mathrm{SS}}\left(r_{\mathrm{A}}\right)+H_0$, where $a$ and $H_0$ are constants. As it is given explicitly in \citet{Erkut2016}, the expression for $H_{\mathrm{SS}}\left(r_{\mathrm{A}}\right)$ includes, in general, the angular momentum efficiency constant, $C$, of order unity. In Figure~\ref{fig4}, we display the dependence of $\delta$ on $\dot{M}$ (solid curve) when the disk is in steady equilibrium. As an illustrative example, Figure~\ref{fig4} is obtained for a neutron star of mass $M=2M_\odot$, radius $R=10$~km, and surface magnetic dipole field strength $B=3\times10^8$~G. In accordance with the $\dot{M}$-range of a typical atoll source, the model source in Figure~\ref{fig4} evolves between $4\times 10^{16}\, \mathrm{g\,s}^{-1}$ and $2.6\times 10^{17}\, \mathrm{g\,s}^{-1}$ to produce kHz QPOs. In the long-term X-ray flux evolution of the system, the source would follow the steady-state trajectory of $\delta$ (solid curve in Figure~\ref{fig4}) if the variation in $\dot{M}$ were sufficiently gradual. In a more realistic case, however, the change in $\dot{M}$ may not be slow enough for the system to establish a steady boundary region with a well-defined $\delta$. The system is then expected to jump from one quasi-steady state to another, i.e., from one fluctuating $\delta$ to another (see the dark spots in Figure~\ref{fig4}). On long time scales (more than a day), it is highly likely that the source undergoes rapid variations in $\dot{M}$ whereas on short time scales (hours), $\dot{M}$ usually changes sufficiently slowly for the source to remain stuck or fluctuate near the same $\delta$. Each discrete $\delta$-neighborhood then defines a particular track in the $\nu_1$ versus $\dot{M}$ plane as shown in Figure~\ref{fig5}. The thin solid curves labeled with a, b, and c stand for the run of lower kHz QPO frequency estimated by the model function in Equation~(\ref{nu1}) when $\delta$ follows its steady-state trajectory (solid curve in Figure~\ref{fig4}) for $\Delta \nu=290$, 350, and 400~Hz, respectively. In Figure~\ref{fig5}, the parallel tracks are generated along each curve representing a different value of $\Delta \nu$ with the use of the same model function evaluated at $\delta$-values fluctuating within the dark patches shown in Figure~\ref{fig4}. Note that the parallel tracks in Figure~\ref{fig5} are reminiscent of those observed in 4U~1608--52 \citep{Mendez2000} and Aql~X-1 \citep{Barret2008}. It is noteworthy that there is a general tendency for the slopes of the tracks to decrease as the source evolves toward higher $\dot{M}$ or intensity state. The saturation effect of ever increasing $\delta$ for sufficiently low values of $\dot{M}$ on $\nu_1$ can also be seen from Figure~\ref{fig5}. As we shall deduce in the next section from the application of our model to the case of observed parallel tracks, the same effect can be explained by the relatively dense distribution of quasi-steady states occupied by the source on the left branch of $\delta$-curve.

\subsection{Model Application to Observed Parallel Tracks}\label{opt}

\begin{figure}
\epsscale{1.16} \plotone{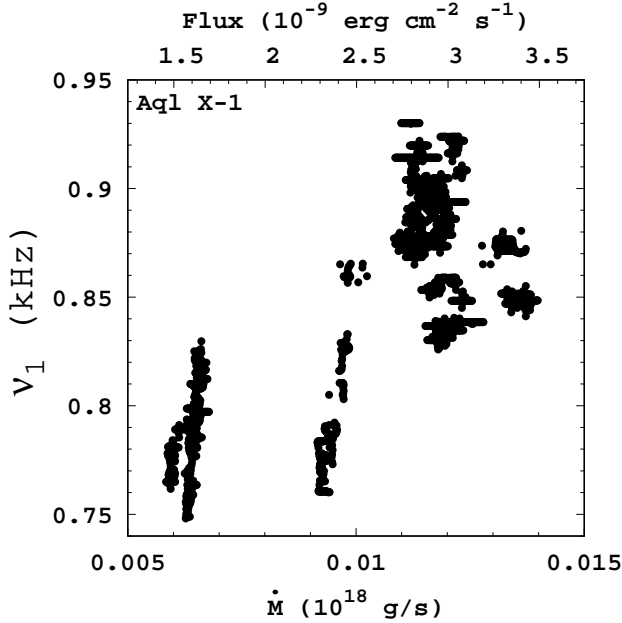} \caption{Lower kHz QPO frequency versus X-ray flux (upper horizontal axis) estimated through comparison of the frequency versus count rate data in \citet{Mendez2000} with those in \citet{Zhang1998} for Aql~X-1. The lower horizontal axis stands for the mass accretion rate, which we deduce from the X-ray flux of the source for a neutron star of $M=1.4M_\odot$ and $R=10$~km using $\dot{M}=4\pi d^2F_{\mathrm{X}}R/GM$ and a source distance of $d=2.5$~kpc \citep{CILP99}. \label{fig6}}
\end{figure}

We consider the parallel tracks observed in the plane of lower kHz QPO frequency versus X-ray count rate or X-ray flux, $F_{\mathrm{X}}$, of 4U~1608--52, 4U~1636--53, and Aql~X-1. The correlation between $\nu_1$ and $F_{\mathrm{X}}$ in the $2-10$~keV range obtained by \citet{Zhang1998} for Aql~X-1 is almost the same as the correlation between $\nu_1$ and the count rate in the $2-16$~keV energy band shown by \citet{Mendez2000} for the same source. The high similarity between the two relations can be seen from the direct comparison of the parallel tracks at the lowest flux values $(1.3-2.5\times 10^{-9}\, \mathrm{ergs}\, \mathrm{cm}^{-2}\, \mathrm{s}^{-1})$ in \citet{Zhang1998} with those at the lowest count rates $(600-1100\, \mathrm{counts}\, \mathrm{s}^{-1})$ in \citet{Mendez2000}. Extending the approximate flux-count rate proportionality to higher count rates as well, we show, in Figure~\ref{fig6}, our estimate for the correlation between $\nu_1$ and $F_{\mathrm{X}}$ in Aql~X-1. We use the frequency versus count rate data in \citet{Mendez2000} in comparison with those in \citet{Zhang1998} to obtain an approximate plot of the parallel tracks in the $\nu_1$ versus $F_{\mathrm{X}}$ plane. In Figure~\ref{fig6}, the upper horizontal axis represents $F_{\mathrm{X}}$. On the lower horizontal axis of the same figure, we show, for a neutron star of mass $M=1.4M_\odot$ and radius $R=10$~km, the mass accretion rate corresponding to the X-ray luminosity, $L_{\mathrm{X}}=4\pi d^2F_{\mathrm{X}}$, where $d$ is the source distance. As in \citet{Erkut2016}, we use $L_{\mathrm{X}}=GM\dot{M}/R$ to the extent that $L_{\mathrm{X}}$ is a good indicator of the accretion luminosity, which seems to be a reasonable assumption in particular for most of the neutron-star LMXBs when the bolometric flux can be approximated by the total flux deduced from the source spectra, which are simultaneously obtained with the QPO frequencies \citep{Ford2000}.

\begin{figure}
\epsscale{1.2} \plotone{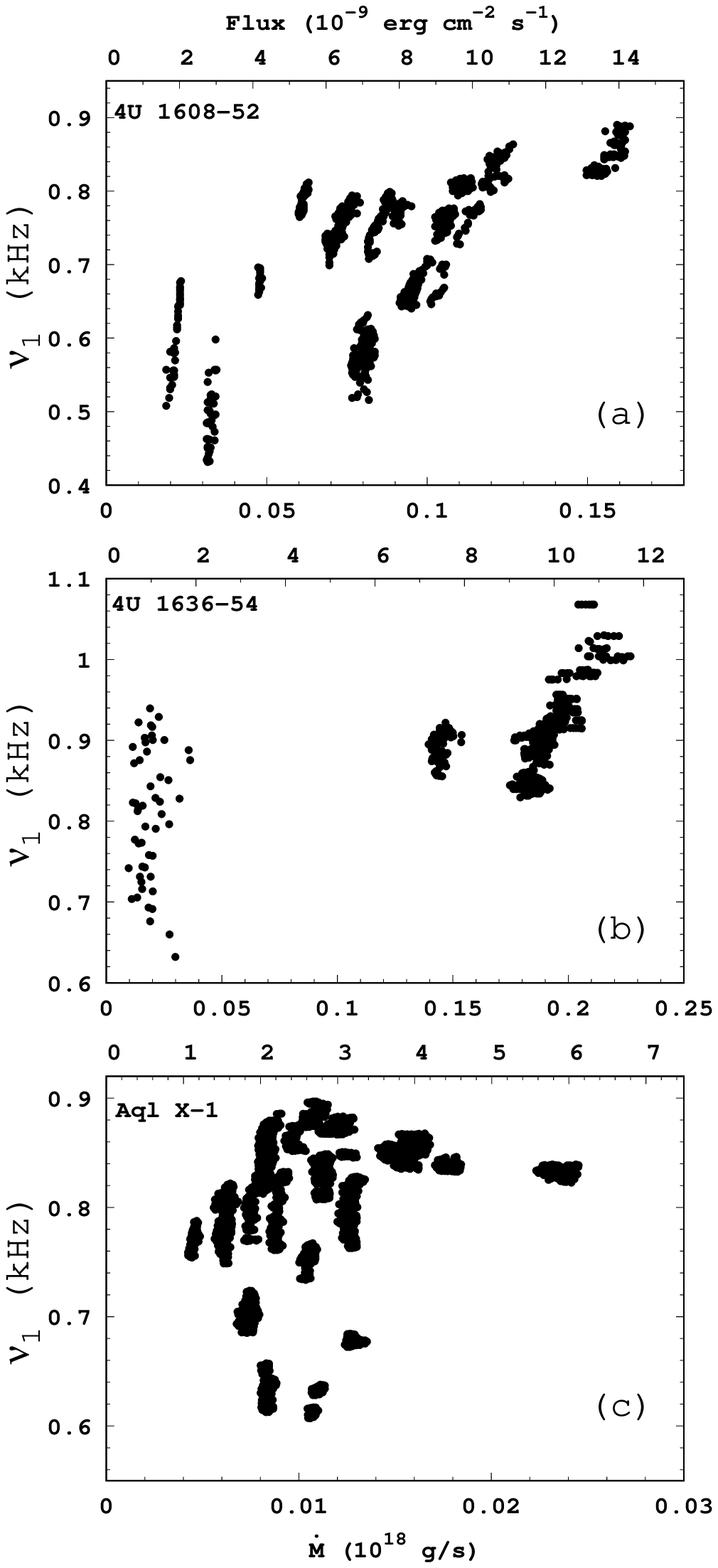} \caption{Lower kHz QPO frequency versus X-ray flux (upper horizontal axis of each panel) estimated through comparison of the frequency versus count rate data \citep{Mendez2000,Mendez2001} with the frequency versus $L_{\mathrm{X}}$ data in \citet{Ford2000} for 4U~1608--52 (panel~(a)), 4U~1636--53 (panel~(b)), and Aql~X-1 (panel~(c)). Panel~(b) includes the relatively recent data of 4U~1636--53 \citep{Belloni2007} when the source is in the low flux regime. In panel~(c), we display the data in \citet{Barret2008}, which comprise also the data in \citet{Mendez2001}. The lower horizontal axis of each panel represents the mass accretion rate, which we deduce from the X-ray flux of the source for a neutron star of $M=1.4M_\odot$ and $R=10$~km using a source distance of $d=4.2$~kpc \citep{Erkut2016} for 4U~1608--52, $d=5.5$~kpc \citep{PW95} for 4U~1636--53, and $d=2.5$~kpc \citep{CILP99} for Aql~X-1. \label{fig7}}
\end{figure}

The $F_{\mathrm{X}}$-range for the parallel tracks of both Aql~X-1 and other sources such as 4U~1608--52 and 4U~1636--53 in \citet{Mendez2000} can also be estimated from the $L_{\mathrm{X}}$-range these sources cover in \citet{Ford2000} with the same frequency range for $\nu_1$. In Figure~\ref{fig7}, we present our estimate for the parallel tracks of individual sources such as 4U~1608--52, 4U~1636--53, and Aql~X-1 in the plane of $\nu_1$ versus $F_{\mathrm{X}}$ (upper horizontal axis in each panel). Taking into account the relatively recent data of 4U~1636--53 \citep{Belloni2007} and using the data in \citet{Barret2008} for Aql~X-1, we consider the long-term flux evolution of the parallel tracks. The plots in Figure~\ref{fig7} can only be regarded as plausible approximations to the exact $\nu_1$ versus $F_{\mathrm{X}}$ relations of these sources, and yet they serve our purpose as samples for studying the lack of correlation between $\nu_{\mathrm{kHz}}$ and $F_{\mathrm{X}}$ of an individual source on long timescales. The lower horizontal axis in each panel of Figure~\ref{fig7} shows the $\dot{M}$-range the source covers according to the $F_{\mathrm{X}}$-range for a chosen pair of $M$ and $R$ of the neutron star and an adopted distance to the source.

\begin{figure*}
\epsscale{1} \plotone{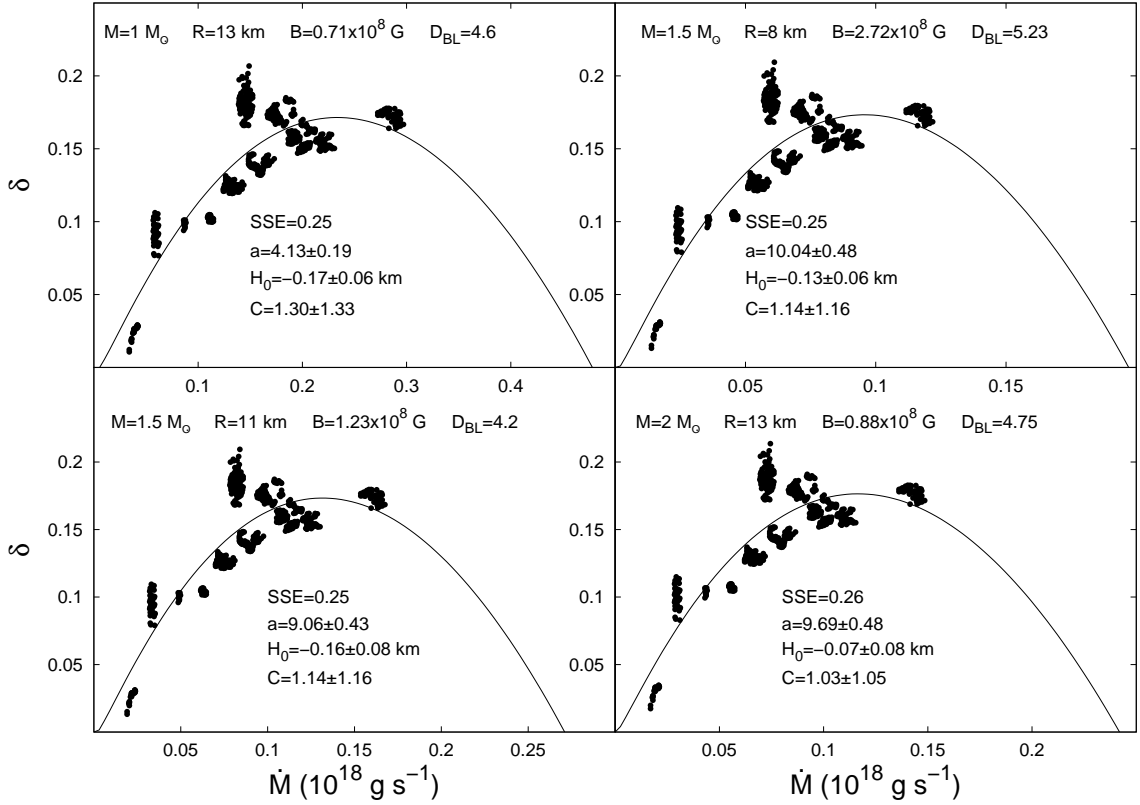} \caption{Numerical data (dark spots) for $\delta$ as a function of $\dot{M}$, which can reproduce the parallel tracks of 4U~1608--52 (Figure~\ref{fig7}) for different values of neutron-star $M$, $R$, and $B$. In all panels, $\Delta \nu=300$~Hz. Solid curve represents the best fit to the numerical data for $\delta$ and acts therefore as an approximate model for the steady state behavior of $\delta$. \label{fig8}}
\end{figure*}

The application of our model to the parallel tracks in Figure~\ref{fig7} is not only limited to certain values of the neutron-star masses and radii, which define the $\dot{M}$-range for a given source distance. In addition to the parameters such as $\Delta \nu$ and $\delta$, which in turn varies with $\dot{M}$, the model function in Equation~(\ref{nu1}) depends on $M$, $R$, $B$, and $\dot{M}$ as well via $\nu_{\mathrm{A}}$ \citep{Erkut2016}. In order to reveal how
$\delta$ is supposed to change for the model function to reproduce the parallel tracks in Figure~\ref{fig7}, we numerically scan $M$, $R$, and $B$ values for which the $\nu_1$ data can be obtained at the corresponding $\dot{M}$-values, first keeping $\Delta \nu$ constant for each source. We then repeat the same procedure for 4U~1608--52 and 4U~1636--53 when $\Delta \nu$ is allowed to vary with $\nu_1$ in accordance with the observations of these two sources \citep{Mendez98b,Jonker2002}. In Figure~\ref{fig7}, the $\dot{M}$-range for each source is based on the adopted distance to the source. The results of our analysis, however, do not change considerably if we adopt a different value for the source distance. We have repeated the numerical scan of $M$, $R$, and $B$ for different source distances available in the literature and obtained similar results.

In Figure~\ref{fig8}, we show the numerical data (dark spots) for $\delta$ as a function of $\dot{M}$ needed to reproduce the parallel tracks of 4U~1608--52 (Figure~\ref{fig7}) for different values of $M$, $R$, and $B$. To obtain Figure~\ref{fig8}, we assume $\Delta \nu=300$~Hz. In each panel of Figure~\ref{fig8}, the generic model for the steady-state $\delta$ is estimated by a thin solid curve that fits the numerical data best. We construct the steady-state model of $\delta$ by referring to the standard disk theory as explained in Section~\ref{mifpt} (see also Figure~\ref{fig4}). The sum of squared errors (SSE) between the values of the fit function and data is minimized using the Marquardt-Levenberg algorithm \citep{Erkut2016}. The nonlinear least squares fitting of the generic model function for $\delta$ also yields the numerical values of the model parameters such as $a$, $H_0$, and $C$. Being consistent with thin disk assumption, we find $D_{\mathrm{BL}}$ by choosing the minimum value of this parameter such that $\varepsilon$ remains always less than a certain limit $(\varepsilon<0.1)$ for the maximum value of $\delta$. The fit examples in Figure~\ref{fig8} are the best fits with smallest SSE and physically plausible values, e.g., $a>0$ (Section~\ref{mifpt}), for the fit parameters among different sets of $M$, $R$, and $B$-values.

\begin{figure*}
\epsscale{1} \plotone{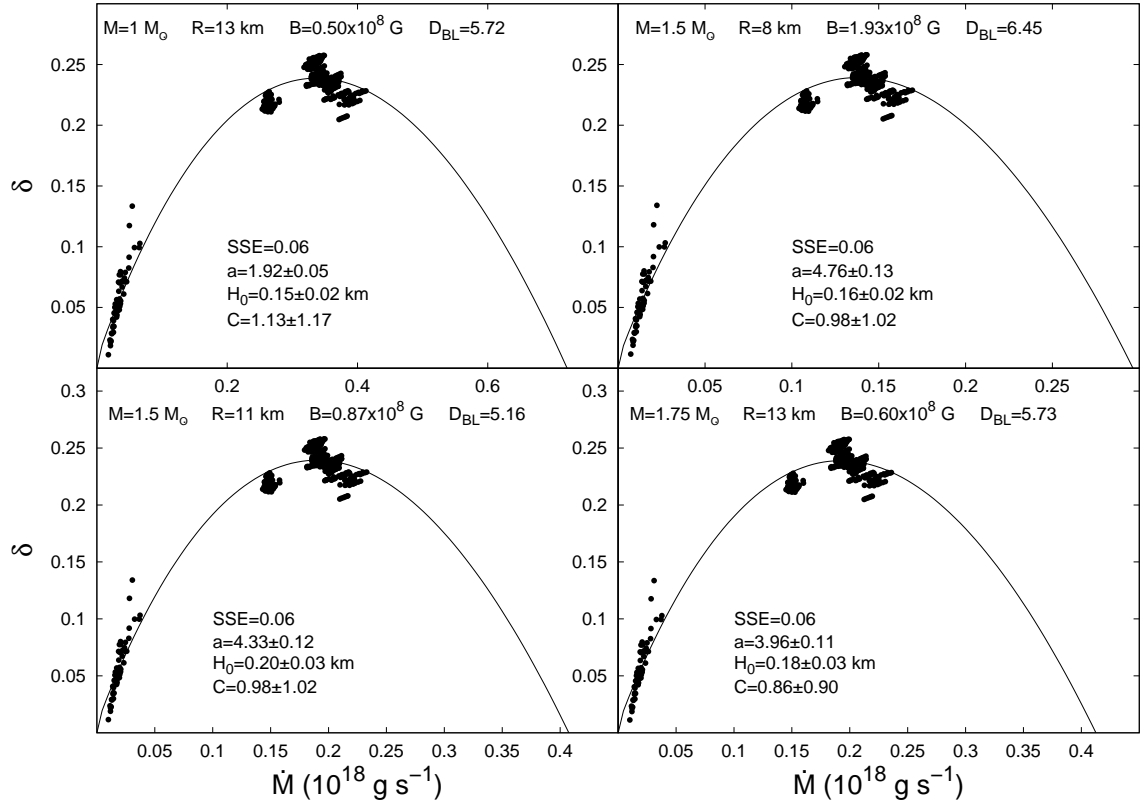} \caption{Same as Figure~\ref{fig8}, but for the regeneration of the parallel tracks of 4U~1636--53 (Figure~\ref{fig7}). \label{fig9}}
\end{figure*}

\begin{figure*}
\epsscale{1} \plotone{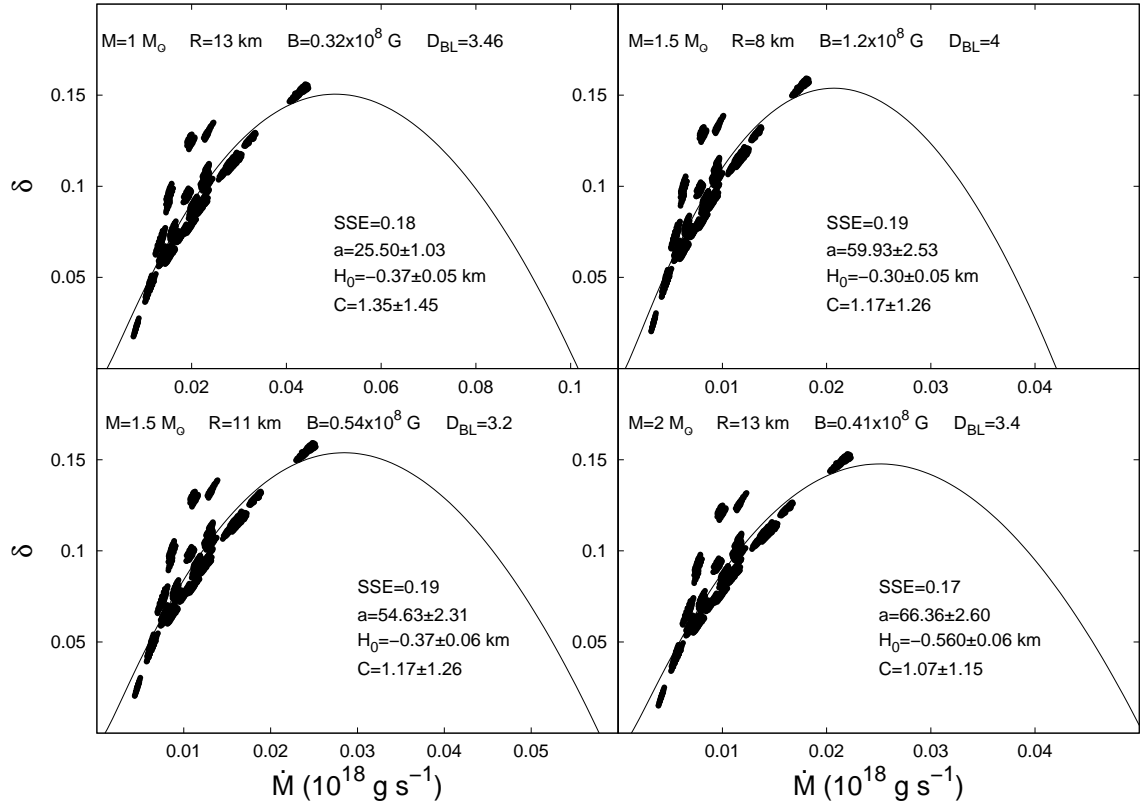} \caption{Same as Figure~\ref{fig8}, but for the regeneration of the parallel tracks of Aql~X-1 (Figure~\ref{fig7}). \label{fig10}}
\end{figure*}

The parallel tracks of 4U~1636--53 and Aql~X-1 (Figure~\ref{fig7}) can be regenerated if $\delta$ varies with $\dot{M}$ as in Figures~\ref{fig9} and \ref{fig10}, respectively. As in the case of 4U~1608--52, the panels with different $M$, $R$, and $B$-values in Figures~\ref{fig9} and \ref{fig10} display the numerical data for the variation of $\delta$ to which we find the best fits using the generic model function (solid curves). It is noteworthy that the numerical data for $\delta$ in all cases (Figures~\ref{fig8}, \ref{fig9}, and \ref{fig10}) seem to follow, albeit with some data scattering, the estimated steady behavior of $\delta$ in the long-term $\dot{M}$ (or flux) evolution of the source. The observed scattering of the numerical data for $\delta$ around the mean trend curve can be due to several effects, which we do not reckon with in our model. These effects might originate from the time-dependent fluctuations of the model parameters such as $D_{\mathrm{BL}}$ and $\Delta \nu$, which we treat as constants at the first stage of our analysis. Next, we address the observed variation of $\Delta \nu$ as $\nu_1$ drifts from lower frequencies to higher frequencies and repeat the same analysis with a variable $\Delta \nu$ instead of keeping it constant at 300~Hz.

\begin{figure*}
\epsscale{1.16} \plottwo{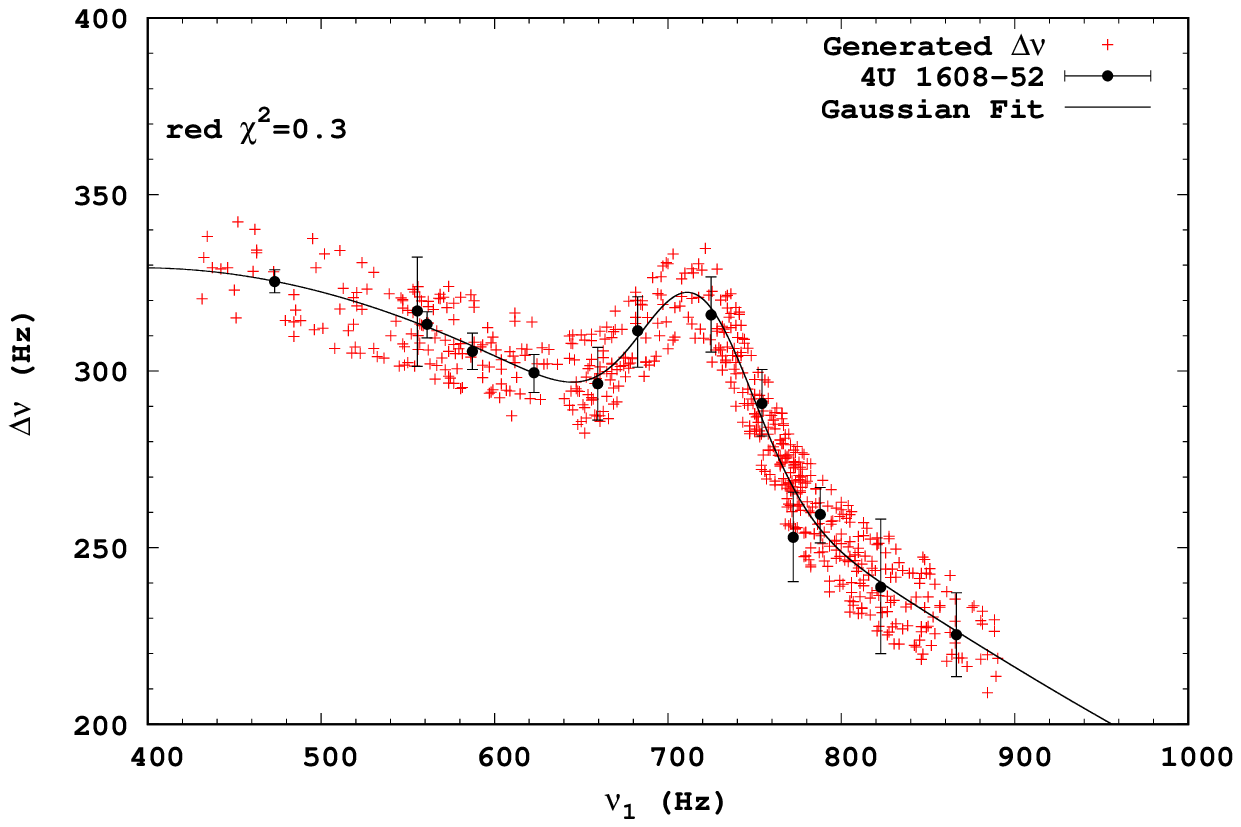}{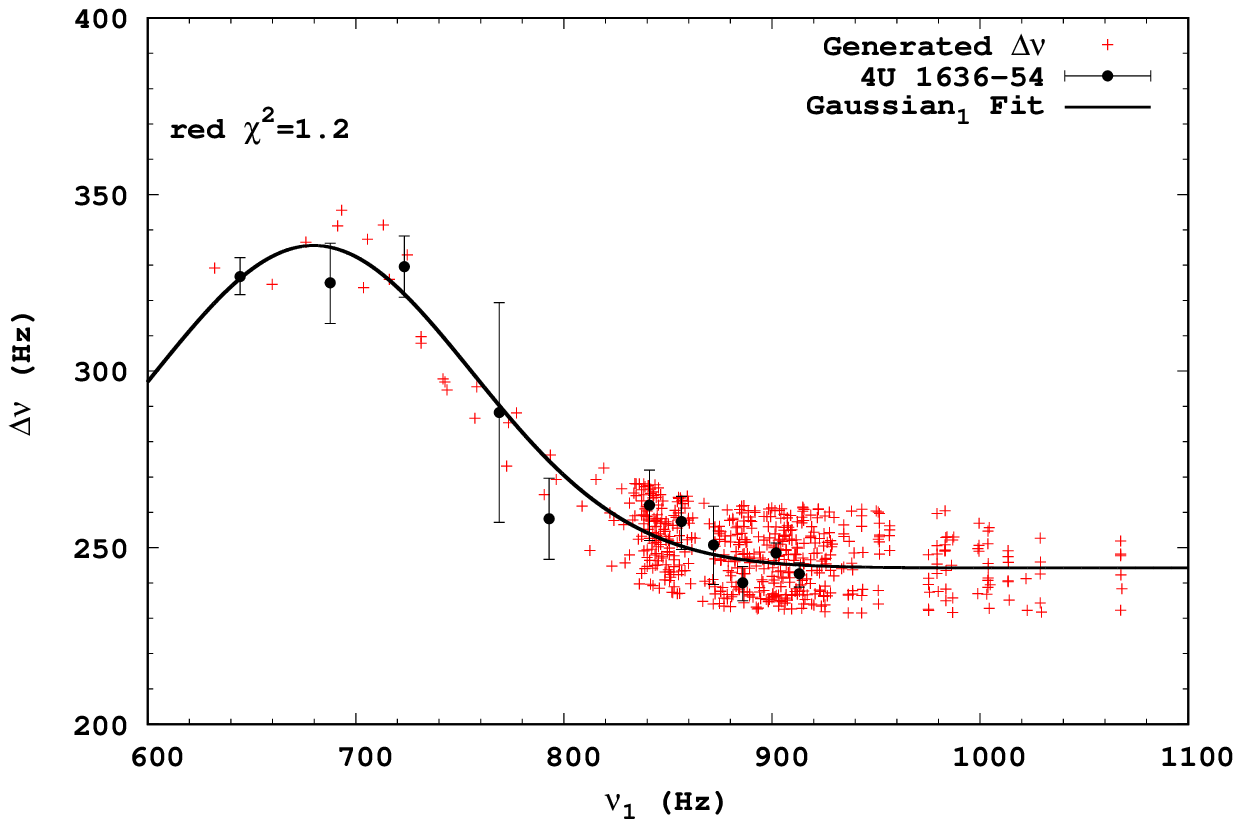} \caption{The frequency difference, $\Delta \nu$, between the upper and lower kHz QPO peaks as a function of the lower kHz QPO frequency, $\nu_1$. The solid curves on the left and right panels represent the best fits to the observational data of 4U~1608--52 \citep{Mendez98b} and 4U~1636--53 \citep{Jonker2002}, respectively. The fit to the data of 4U~1608--52 is fairly described by two Gaussian functions, which yield a reduced $\chi^2$ of $\sim 0.3$. While a single Gaussian function cannot account for the data of 4U~1608--52, it fits the data of 4U~1636--53 quite well with a reduced $\chi^2$ of $\sim 1.2$. The numerical data (red plus signs) for $\Delta \nu$ are randomly generated within an interval of $\sim 30$~Hz around the fit curve to reproduce the parallel tracks in Figure~\ref{fig7}. \label{fig11}}
\end{figure*}

\begin{figure*}
\epsscale{1} \plotone{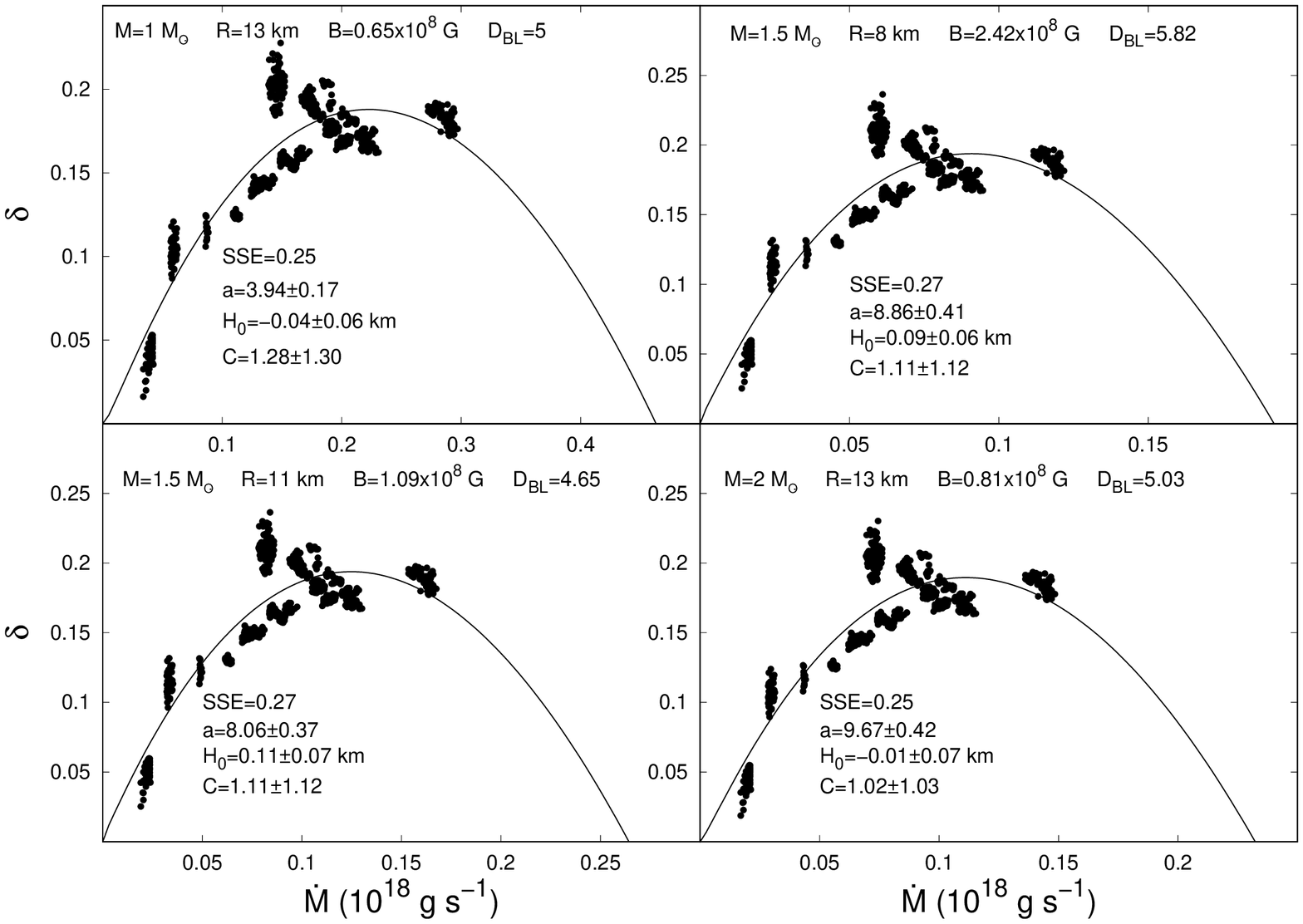} \caption{Numerical data (dark spots) for $\delta$ as a function of $\dot{M}$, which can reproduce the parallel tracks of 4U~1608--52 (Figure~\ref{fig7}) for different values of neutron-star $M$, $R$, and $B$. In all panels, $\Delta \nu$ is extracted from the randomly generated data (red plus signs) around the fit curve in the left panel of Figure~\ref{fig11}. Solid curve represents the best fit to the numerical data for $\delta$ and acts therefore as an approximate model for the steady state-behavior of $\delta$. \label{fig12}}
\end{figure*}

\begin{figure*}
\epsscale{1} \plotone{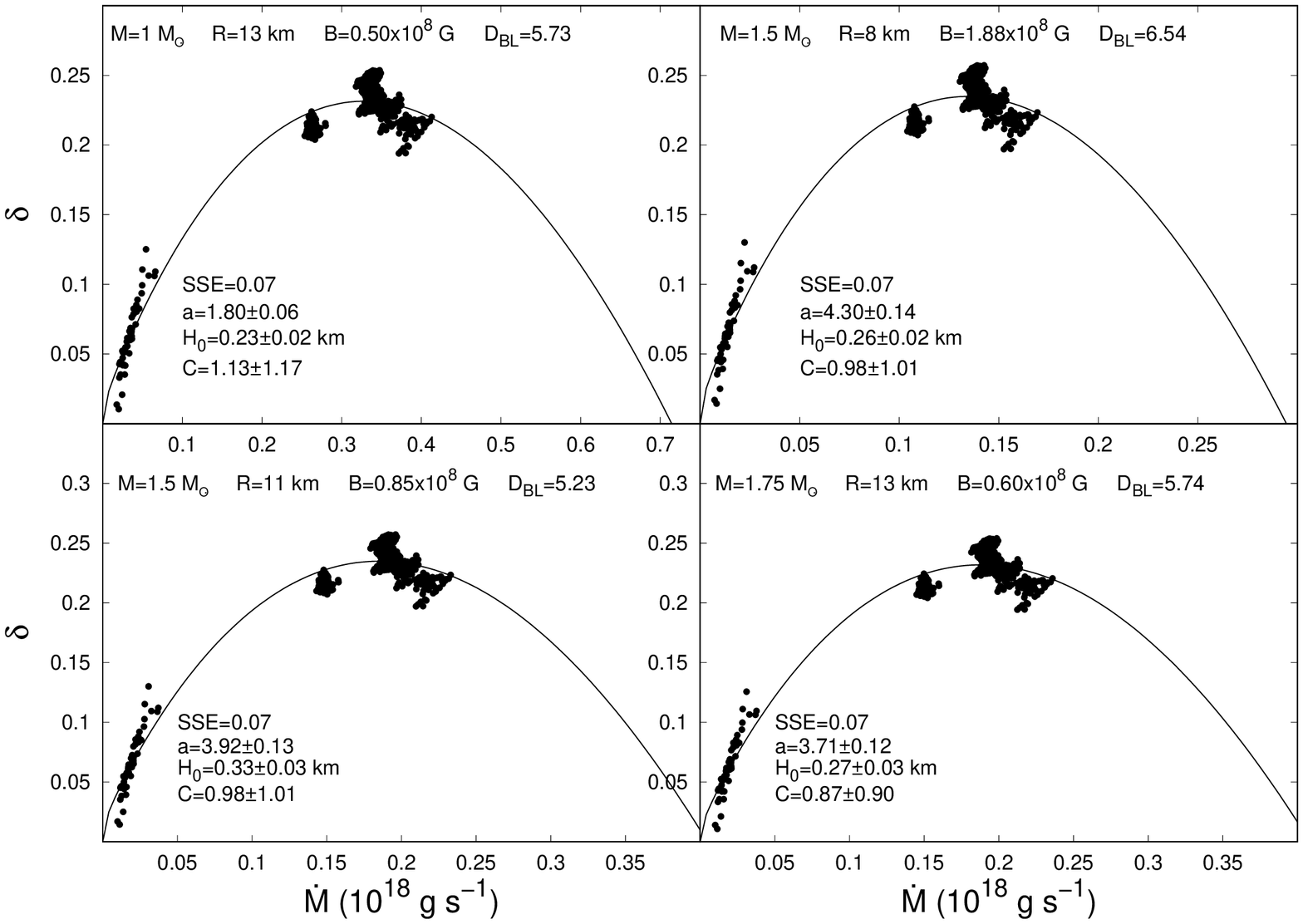} \caption{Numerical data (dark spots) for $\delta$ as a function of $\dot{M}$, which can reproduce the parallel tracks of 4U~1636--53 (Figure~\ref{fig7}) for different values of neutron-star $M$, $R$, and $B$. In all panels, $\Delta \nu$ is extracted from the randomly generated data (red plus signs) around the fit curve in the right panel of Figure~\ref{fig11}. Solid curve represents the best fit to the numerical data for $\delta$ and acts therefore as an approximate model for the steady-state behavior of $\delta$. \label{fig13}}
\end{figure*}

Unlike Aql~X-1, where only one kHz QPO has been reported except in one case \citep{Barret2008}, two kHz QPO peaks have usually been detected in the power spectra of 4U~1608--52 and 4U~1636--53. The peak separation $\Delta \nu$ between the two kHz QPO frequencies has been measured using the average spectra of data sets consisting of individual power spectra aligned with respect to $\nu_1$ \citep{Mendez98b}. The so-obtained plots of $\Delta \nu$ versus $\nu_1$ can therefore describe the average trend in the correlation between these two quantities. The variation of $\Delta \nu$ with $\nu_1$, within different parallel tracks including the ones with similar $\nu_1$-range, however, may not be identical. Tracks with overlapping ranges for $\nu_1$ might correspond to non-overlapping ranges for $\Delta \nu$. Yet, the average trend in the correlation between $\Delta \nu$ and $\nu_1$ can be used to generate a set of $\Delta \nu$-values changing with $\nu_1$, which we substitute into the model function (Equation~(\ref{nu1})) to see how $\delta$ varies numerically with $\dot{M}$ for the regeneration of parallel tracks in Figure~\ref{fig7}.

The left and right panels of Figure~\ref{fig11} manifest the best fits (solid curves) to the observational data for $\Delta \nu$ as a function of $\nu_1$ in 4U~1608--52 and 4U~1636--53, respectively. We use the randomly generated $\Delta \nu$-values (red plus signs) around the fit curve as a data sample to reproduce the parallel tracks of a given source. As it is seen from Figure~\ref{fig12}, the numerical data for $\delta$ as a function of $\dot{M}$ to reproduce the parallel tracks of 4U~1608--52 are very similar to those in Figure~\ref{fig8}. The same conclusion holds if we compare Figure~\ref{fig13}, where $\Delta \nu$ evolves according to the average tendency in the frequency-frequency correlation, with Figure~\ref{fig9}, where $\Delta \nu$ is kept constant. The distribution of the numerical data for $\delta$ does not change considerably when $\Delta \nu$ varies slowly around the average course of frequency-frequency correlation (Figure~\ref{fig11}) in comparison with the data distribution when $\Delta \nu$ is constant. One of the reasons why scattering of $\delta$-data around the steady state is relatively more pronounced for some of the tracks (each parallel track corresponds to a dark spot in the $\delta$-$\dot{M}$ plot) could be the sufficiently large difference between our choice of $\Delta \nu$ (either being constant or changing around an average value) and the set of actual $\Delta \nu$-values for that particular track. The effect of sufficiently large differences ($\gtrsim 100$~Hz) between the values of $\Delta \nu$ on the distribution of parallel tracks can also be seen from Figure~\ref{fig5}.

The scattering of numerical data around the fit curve in the $\delta$-$\dot{M}$ plane may also originate from the variations in the effective magnetic diffusivity, $\eta$, of the boundary layer. Reconnection between toroidal magnetic field components in opposite directions leads to lower effective conductivity (higher $\eta$) as compared to the usual Coulomb or Bohm conductivity \citep{GL1979}. Being associated with reconnection events, the enhanced $\eta$ in the boundary layer does not only limit the growth of $B^+_{\phi}$, resulting in relatively small values of the azimuthal pitch, $\gamma_{\phi}$, but also the screening of the poloidal magnetic field. The boundary region for the efficient attenuation of the poloidal field gets broadened in the radial direction (see Section~\ref{rdbr}). Both effects can be realized from Equations~(\ref{slnbz}) and (\ref{omgzr}). As the coefficient of magnetic diffusivity, $D_{\mathrm{BL}}$, and therefore $\eta$ increases, the attenuation of $B_z$ becomes less steep while $\left|\gamma_{\phi}\right|$ decreases and vice versa. To study how $\left|\gamma_{\phi}\right|$ evolves as a function of $\dot{M}$, we substitute Equation~(\ref{bc}) into Equation~(\ref{omgzr}) to write
\begin{equation}
\left|\gamma_{\phi}\right|=\frac{2}{D_{\mathrm{BL}}b_0^2}\left(1-\frac{\Delta \nu}{\nu_{\mathrm{A}}}\right), \label{gamafi}
\end{equation}
where $\nu_{\mathrm{A}}=\nu_{\mathrm{K}}\left(r_{\mathrm{A}}\right)\propto \dot{M}^{3/7}$. Assuming that $D_{\mathrm{BL}}$ and $\Delta \nu$ are constants, the $\dot{M}$-dependence of the azimuthal pitch can be estimated as
\begin{equation}
\left|\gamma_{\phi}\right|=b+d\dot{m}^{-3/7}, \label{gama}
\end{equation}
with $b$ and $d$ being positive and negative constants, respectively. Here, $\dot{m}$ is the mass accretion rate in units of $10^{18}\, \mathrm{g}\, \mathrm{s}^{-1}$.

\begin{figure}
\epsscale{1.16} \plotone{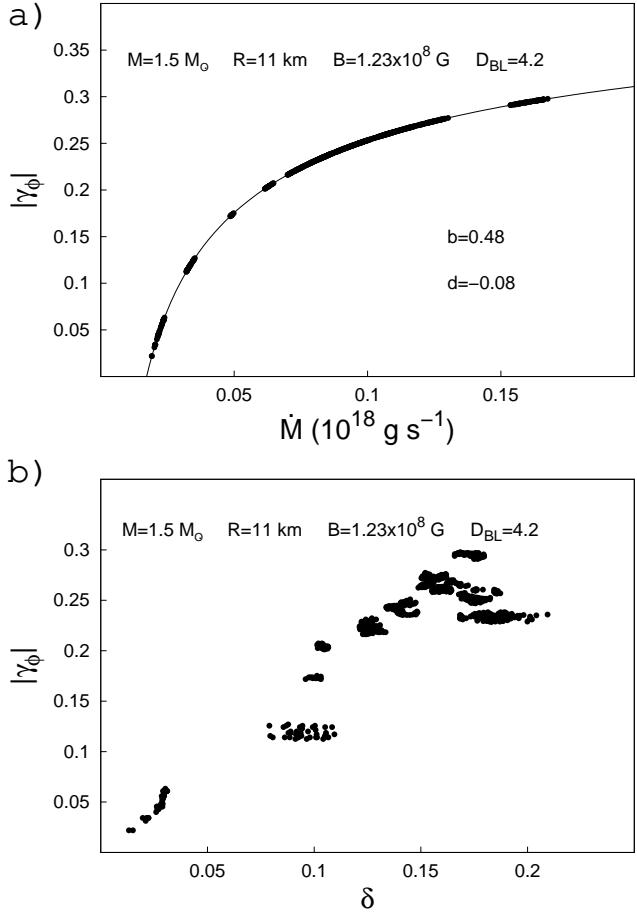} \caption{Azimuthal pitch magnitude, $\left|\gamma_{\phi}\right|$, as a function of $\dot{M}$ according to Equation~(\ref{gama}) for $\left|b_0\right|=1$, $D_{\mathrm{BL}}=4.2$, and $\Delta \nu=300$~Hz for a neutron star of $M=1.5M_\odot$, $R=11$~km, and $B=1.23\times 10^8$~G (solid curve). Black dots on the curve correspond to the values of $\dot{M}$, at which the parallel tracks of 4U~1608--52 can be reproduced (panel~(a)). Relation between $\left|\gamma_{\phi}\right|$ for $D_{\mathrm{BL}}=4.2$ and the numerical $\delta$-data on the bottom left panel of Figure~\ref{fig8} (panel~(b)). \label{fig14}}
\end{figure}

\begin{figure}
\epsscale{1.16} \plotone{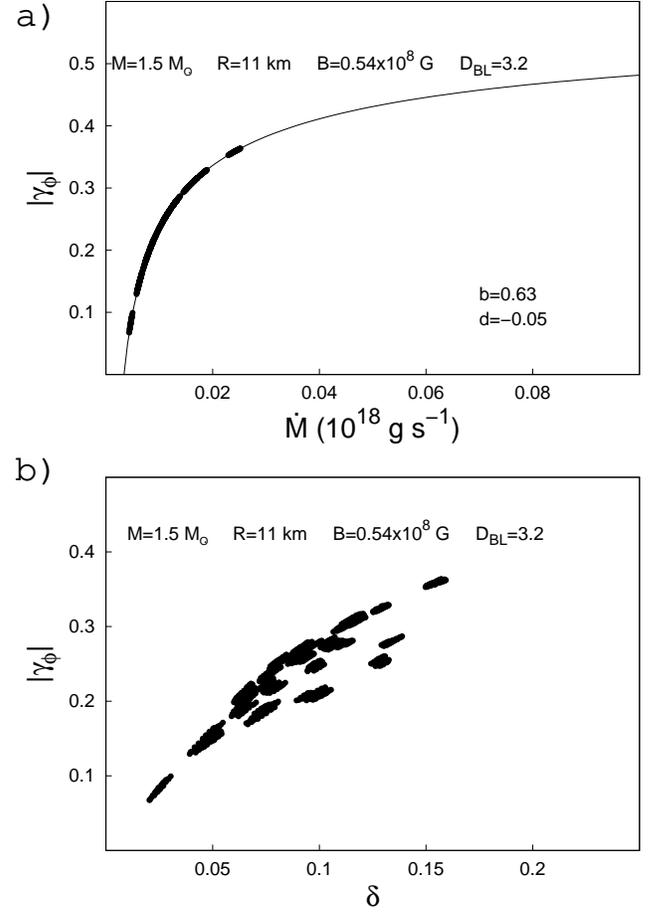} \caption{Azimuthal pitch magnitude, $\left|\gamma_{\phi}\right|$, as a function of $\dot{M}$ according to Equation~(\ref{gama}) for $\left|b_0\right|=1$, $D_{\mathrm{BL}}=3.2$, and $\Delta \nu=300$~Hz for a neutron star of $M=1.5M_\odot$, $R=11$~km, and $B=0.54\times 10^8$~G (solid curve). Black dots on the curve correspond to the values of $\dot{M}$, at which the parallel tracks of Aql~X-1 can be reproduced (panel~(a)). Relation between $\left|\gamma_{\phi}\right|$ for $D_{\mathrm{BL}}=3.2$ and the numerical $\delta$-data on the bottom left panel of Figure~\ref{fig10} (panel~(b)). \label{fig15}}
\end{figure}

In Figure~\ref{fig14}, the panel~(a) displays the run of $\left|\gamma_{\phi}\right|$ as $\dot{M}$ changes in accordance with Equation~(\ref{gama}) provided $\left|b_0\right|=1$, $D_{\mathrm{BL}}=4.2$, and $\Delta \nu=300$~Hz for a neutron star of $M=1.5M_\odot$, $R=11$~km, and $B=1.23\times 10^8$~G. The black dots on the solid curve correspond to the values of $\dot{M}$, at which the parallel tracks of 4U~1608--52 can be reproduced for the same values of $M$, $R$, and $B$ by the numerical data for $\delta$ in Figure~\ref{fig8}. Note that the solid curves for $\left|\gamma_{\phi}\right|$ (panel~(a) of Figure~\ref{fig14}) and $\delta$ (bottom left panel of Figure~\ref{fig8}) are plotted for the same constant value of $D_{\mathrm{BL}}$, i.e., for $D_{\mathrm{BL}}=4.2$, and therefore describe the same average steady-state behavior. If all numerical data for $\delta$ in Figure~\ref{fig8} appeared to be arrayed along the steady-state curve for the given $D_{\mathrm{BL}}$, we would expect to find a strong correlation between the $\left|\gamma_{\phi}\right|$-values for the same $D_{\mathrm{BL}}$ and the numerical data for $\delta$, which are supposed to regenerate the parallel tracks. The panel~(b) of Figure~\ref{fig14} shows the relation between $\left|\gamma_{\phi}\right|$ for $D_{\mathrm{BL}}=4.2$ and the numerical data for $\delta$ (bottom left panel of Figure~\ref{fig8}). We observe that most of the data indicate the existence of a correlation between the two quantities in alignment with our expectations. We know from their steady-state behavior for the same $D_{\mathrm{BL}}$ that both $\left|\gamma_{\phi}\right|$ and $\delta$ increase with $\dot{M}$. The data clusters, which are centered around the coordinates $(\delta=0.1,\left|\gamma_{\phi}\right|=0.12)$ and $(\delta=0.19,\left|\gamma_{\phi}\right|=0.23)$, appear to vitiate the general trend in the $\left|\gamma_{\phi}\right|$-$\delta$ relation by attaining significantly higher $\delta$-values as compared to others (panel~(b) of Figure~\ref{fig14}). Being examples of the most prominent deviation from the steady state labeled with $D_{\mathrm{BL}}=4.2$, these two data clusters can be identified with the two dark spots lying well above the solid curve in Figure~\ref{fig8}. We would reach similar conclusions if we repeated the same analysis for different sources or the same source with different values of $M$, $R$, and $B$.

Another good example to study the reason for data scattering around a reference steady state is Aql~X-1. The panels~(a) and (b) of Figure~\ref{fig15} are analogous to their counterparts in Figure~\ref{fig14} with the use of the same neutron-star mass and radius. The relatively high $\delta$-values associated with the data clusters having coordinates $(\delta=0.1,\left|\gamma_{\phi}\right|\simeq0.2)$ and $(\delta \simeq0.14,\left|\gamma_{\phi}\right|\simeq0.25)$ as compared to others, which seem to follow more or less the average steady state labeled with $D_{\mathrm{BL}}=3.2$, is the main reason for the data scattering in the $\delta$-$\dot{M}$ plane (bottom left panel of Figure~\ref{fig10}). Note that both $\delta=\varepsilon D_{\mathrm{BL}}/2$ and $\left|\gamma_{\phi}\right|$ (Equation~(\ref{gamafi})) depend on $D_{\mathrm{BL}}$. Any $\delta$-data points falling in a range of values, which are incompatible with the $\left|\gamma_{\phi}\right|$-$\delta$ relation, must therefore be in a different steady or quasi-steady state. Data clusters characterized by relatively higher values of $\delta$ (Figures~\ref{fig14} and \ref{fig15}) are expected to be labeled with larger $D_{\mathrm{BL}}$-values in comparison with that of the average steady state and vice versa. The existence of quasi-steady states the source would occupy with certain $D_{\mathrm{BL}}$-values deviating from the average (or reference) value is highly likely due to magnetic reconnection events taking place in the boundary region.

\section{Discussion and Conclusions}\label{disc}

The recent study by \citet{Erkut2016} to explain the lack of correlation between $\nu_{\mathrm{kHz}}$ and $L_{\mathrm{X}}$ in the ensemble of neutron-star LMXBs has revealed the existence of a possible correlation between $\nu_1$ and $\dot{M}/B^2$. The distribution of QPO frequencies in the $\nu_1$-$\dot{M}/B^2$ plane can be accounted for by the cumulative effect of the model function fit to individual source data. In this paper, we calculate $\nu_1$ using a model function (Equation~(\ref{nu1})), which takes into account the exact matching of the non-Keplerian boundary region with the Keplerian disk. In comparison with the model function employed in \citet{Erkut2016}, the present one estimates lower $\nu_1$-values for the same $\delta$ at a given $\dot{M}$. The region spanned by the present model function in the $\nu_1$-$\dot{M}/B^2$ plane (or $\nu_1$-$\dot{M}$ plane for a given $B$) is relatively broader for the same range of $\delta$ (Figure~\ref{fig1}). The observed distribution of $\nu_{\mathrm{kHz}}$ in both individual sources and their ensemble can therefore be accounted for within a narrower range for $\delta$ using the function in Equation~(\ref{nu1}). Although they are slightly different from each other as far as the $\delta$-range is concerned, the model functions in \citet{Erkut2016} and the present work have the same dependence on $\dot{M}$, i.e., the slopes of both functions decrease as $\delta$ increases and vice versa (Figure~\ref{fig1}). Instead of employing the function in Equation~(\ref{nu1}) for $\nu_1$, we could as well use the model function in \citet{Erkut2016} to generate parallel tracks. We would find similar results, but for slightly higher values of $\delta$. The model function we derive in Section~\ref{theory}, however, depends on the well defined parameters such as $\delta$ and $\Delta \nu/\nu_{\mathrm{A}}$, which are given in terms of $\varepsilon$ (typical aspect ratio of the disk) and $\gamma_{\phi}$ (azimuthal pitch in the boundary region) (Equations~(\ref{epsilon}), (\ref{omgzr}), and (\ref{bc})), and therefore has the advantage of unveiling the physical mechanism behind parallel tracks within theory of magnetosphere-disk interaction.

\subsection{Summary of Analysis}\label{sum}

The variations of the model parameters such as $\Delta \nu$ and $\delta$ have several important implications for parallel tracks. If both of these parameters are kept constant, the model function in Equation~(\ref{nu1}) cannot produce any parallel tracks. This is because all frequencies would be lined up along a single curve labeled with the constant values of $\delta$ and $\Delta \nu$ in the $\nu_1$ versus $\dot{M}$ plane. Keeping $\delta$ constant, however, changing $\Delta \nu$ on the same plane may lead to two different types of tracks, one with a positive slope and the other with a negative slope, as shown in Figure~\ref{fig2}. As in the case of Sco~X-1, the tracks with negative slopes may arise for sufficiently small values of $\delta$ when $\Delta \nu$ decreases with increasing $\nu_1$ in accordance with the observed correlation between the two kHz QPO frequencies \citep{Psaltis1998}. The tracks labeled with relatively small $\delta$-values are characterized by larger slopes and thus by narrower ranges for $\dot{M}$-variations as compared to the tracks with larger $\delta$, for which $\Delta \nu$ obeys the same frequency-frequency correlation. Keeping $\Delta \nu$ constant while modifying $\delta$ as in Figure~\ref{fig3} results in parallel tracks with positive slopes. In particular, we expect from the morphology and distribution of the tracks observed in atoll sources that $\delta$ increases with the X-ray intensity during the long-term flux evolution of the source, not only because this parameter is supposed to vary with $\dot{M}$ in alignment with the theory of disk accretion, but also because the kHz QPO frequency range of low-intensity tracks is larger than that of high-intensity tracks \citep{Mendez2000,MS2004,Barret2008}.

As shown in Section~\ref{rdbr}, $\delta$ is directly proportional to the coefficient of magnetic diffusivity, $D_{\mathrm{BL}}$, in the boundary layer and the typical aspect ratio of the disk,
$\varepsilon$. Modeling the long-term steady-state variation of $\varepsilon$ with $\dot{M}$ is straightforward if we keep $D_{\mathrm{BL}}$ constant and use the plausible assumption that the typical half-thickness of the disk is in correlation with that of a standard disk at the magnetopause \citep{Erkut2016}. As shown in Figure~\ref{fig4}, we use a typical example of the steady-state behavior of $\delta$ to generate the parallel tracks in Figure~\ref{fig5} by choosing a set of quasi-steady states the source can occupy only at certain values of $\dot{M}$. The source is expected to jump from one state to another unless the variation in $\dot{M}$ is sufficiently slow; otherwise $\delta$ would follow its steady-state trajectory. We obtain the numerical data of each track by evaluating the model function at $\delta$-values fluctuating about the steady solution (solid curve in Figure~\ref{fig4}). Fluctuations correspond to certain $\dot{M}$-intervals being sufficiently narrow, and thus representing relatively short time scales, on which the rapid variation in $\dot{M}$ is highly improbable. In the long-term evolution, we expect the source to seek a quasi-steady state consisting of a set of $\delta$- and $\dot{M}$-values and be stuck there on a short time scale until a sudden variation in $\dot{M}$ occurs. Each parallel track then coincides with one of the possible quasi-steady states the boundary region can hold to produce kHz QPOs.

We present model applications to the observed data of parallel tracks of individual sources such as 4U~1608--52, 4U~1636--53, and Aql~X-1 to disclose the long-term behavior of $\delta$. For both constant and changing $\Delta \nu$, we calculate the numerical data for $\delta$ as a function of $\dot{M}$, which we need to regenerate the parallel tracks. In all three sources, for which the long-term flux evolution can be roughly estimated, the distribution of the numerical $\delta$-data can be successfully fitted by the generic model function for the steady-state $\delta$ in the long term. The model function for $\delta=\varepsilon D_{\mathrm{BL}}/2$ is obtained for a constant value of $D_{\mathrm{BL}}$ and should therefore be regarded as a mean trend describing the average steady behavior of $\delta$. Scattering of numerical data around the fit curve in the plane of $\delta$ versus $\dot{M}$ can be due to several effects, which we do not incorporate into the present form of our model. These effects include the uncertainties  regarding the variations in $\Delta \nu$ and $D_{\mathrm{BL}}$. For some of the tracks, scattering of $\delta$-data around the steady state can be due to large deviation of the actual $\Delta \nu$-values for a particular track from those we use to mimic the observed $\Delta \nu$-$\nu_1$ correlation. Being much more effective than the change in $\Delta \nu$, the variation in $D_{\mathrm{BL}}$ can explain the data scattering in the $\delta$-$\dot{M}$ plane. The reconnection between toroidal magnetic fields in opposite directions may result in enhancement of the effective magnetic diffusivity in the boundary layer \citep{GL1979}. The same process leads to reduction in the magnitude of the azimuthal pitch, $\left|\gamma_{\phi}\right|$, via increase in $D_{\mathrm{BL}}$. We expect $\delta$ and $\left|\gamma_{\phi}\right|$ to be strongly correlated with each other if both parameters evolve nearly around the average steady state labeled with a constant $D_{\mathrm{BL}}$ (Figures~\ref{fig14} and \ref{fig15}). Any data cluster with a set of $\delta$-values violating the correlation between $\left|\gamma_{\phi}\right|$ and $\delta$, which is based on the average steady behavior of these two parameters, define a new quasi-steady state with a different $D_{\mathrm{BL}}$. In general, all numerical $\delta$-data we employ to reproduce the parallel tracks in a given source delineate the possible quasi-steady states the source can occupy in response to variations in the magnetic diffusivity of the boundary region.

\subsection{Lack of Pulsations for Magnetically Channeled Flows}\label{lack}

The presence of quasi-stationary boundary regions whose properties are mainly determined by dynamically important magnetic fields of stellar origin requires, at least partially, the accretion of the disk matter onto the surface of the neutron star via funnels. In analogy with X-ray pulsars in binaries, where strongly magnetized neutron stars accrete mass from their companions, the majority of neutron stars in LMXBs are also expected to pulsate if accretion from the inner disk to the magnetic poles on the stellar surface proceeds through magnetically channeled flows. However, only close to 15\% of neutron stars in LMXBs have so far exhibited coherent pulsations of the persistent flux. Apart from these accretion-powered millisecond oscillations, nuclear-powered millisecond oscillations have also been observed during thermonuclear X-ray bursts. Sources such as SAX~J1808.4--3658 and XTE~J1814--338 have exhibited both accretion- and nuclear-powered X-ray pulses that are similar to one another in pulse shapes and phases. Among accreting millisecond X-ray pulsars (AMXPs), a few of them such as SAX~J1748.9--2021, HETE~J1900.1--2455, and Aql~X-1 are intermittent AMXPs whose pulsations can only be detected sporadically.

The tendency to observe AMXPs mostly in compact transient systems, where the long-term average rate of mass accretion is relatively low as compared to non-pulsating systems might indicate the role of the end product of the magnetic field evolution. As suggested by \citet{Lamb2009}, throughout the spin evolutionary period of the neutron star, spin-up torques acting at high accretion rates would lead to the alignment of the magnetic poles on the surface with the spin axis. As the long-term average of $\dot{M}$ diminishes in the late stage of binary evolution, spin-down torques due to propeller effect or magnetic dipole braking mechanism would cause the neutron-star magnetic poles to move away from the spin axis to render coherent pulsations visible. According to \citet{Lamb2009}, the accretion induced X-ray emitting regions on the stellar surface are likely to be near the spin axis if the star's magnetic field is dynamically important in neutron-star LMXBs. Depending on the motion of emitting regions as $\dot{M}$ varies and the angle between the rotation and magnetic axes, some neutron stars in LMXBs may appear as persistent or intermittent AMXPs whereas others show no pulsations at all.

The X-ray emitting spots that give rise to coherent pulsations of the persistent flux are expected to form when accreting plasma reaches the stellar surface via channeled flow along field lines. For magnetospheric sizes as small as in Figures~\ref{fig8}-\ref{fig13}, where the ratio of the magnetospheric radius to the neutron-star radius changes between 1.3 and 4.6 on average $(r_\mathrm{in}/R\lesssim 5)$ as $\dot{M}$ varies, part of matter may also reach the stellar surface via Rayleigh-Taylor (RT) unstable flow between magnetic field lines and lead to the formation of irregular light curves for small enough values of the angle between magnetic and spin axes \citep{Romanova2008}. The disappearance of coherent pulsations due to multichannel accretion induced by RT instability can therefore be another reason why accretion-powered oscillations are difficult to be resolved. The global three-dimensional hydromagnetic simulations revealed that the high non-axisymmetric modes of RT instability can be suppressed by the presence of the azimuthal field component $B_\phi$ at the inner edge of the accretion disk. It is noteworthy that the range of $\left|\gamma_{\phi}\right|$-values in Figures~\ref{fig14} and \ref{fig15} is in remarkable agreement with the results of these simulations, where $B_\phi$ at $r_\mathrm{in}$ was found to be $\sim (5-30)\%$ of the poloidal field component \citep{Romanova2008}.

\subsection{Magnetic versus Non-Magnetic Boundary Layers}\label{angmombl}

The boundary layer problem involves the rotational dynamics of accretion flow in the innermost disk regions, where the azimuthal velocity of matter becomes sub-Keplerian due to the support of pressure gradients (thermal or magnetic) against gravity in the radial momentum balance while matter accretes due to the efficient transport of angular momentum. The boundary layer we consider in the present work is an electromagnetic boundary layer as in \citet{GL1979}. The dominant stresses that drive the angular momentum transport are due to large scale magnetic fields of stellar origin. Albeit small in size (a few stellar radii), the neutron-star magnetosphere interacts with the innermost disk matter and acts as the source of magnetic stresses that would transfer angular momentum from the boundary-layer plasma to the neutron star (Equation~(\ref{angmom})). In the case of non-magnetic boundary layers around neutron stars with dynamically unimportant magnetic fields, the disk is assumed to extend down to the stellar surface without any interaction with the neutron-star magnetosphere and the rapidly rotating disk matter is envisaged to spread over the neutron-star surface in the form of a thin layer \citep{IS99}. The problem of braking the Keplerian flow using turbulent viscosity of local origin at the base of the spreading layer then requires extremely small values of the viscosity coefficient. Instead of local viscous stresses, giant solitary gravity waves generated by large azimuthal shear may interact with both the upper and lower parts across the layer to transfer angular momentum downward to the neutron-star crust \citep{IS2010}. In the absence of stellar magnetic field, acoustic waves excited by supersonic shear in the boundary layer may also serve as an efficient mechanism of transferring angular momentum, mass, and energy at the disk-star interface \citep{Belyaev13a}. In the magnetohydrodynamic (MHD) simulations by \citet{Belyaev13b}, both the disk and the boundary layer were initially threaded by magnetic fields of different geometry. Magnetosonic waves were found to be important for the angular momentum transport throughout the boundary layer. In all these MHD simulations, however, the accreting star was assumed to have no magnetic field.

The magnetic boundary layer depicted in Section~\ref{theory} is the base of the funnel flow linking the inner disk radius and the surface of the neutron star. The magnetic boundary region is an extension of the innermost disk and therefore can be studied using cylindrical coordinates as in \citet{GL1979}. The deposition of matter onto the stellar surface is expected to occur through the funnel-like flows, which are guided by the magnetic field lines of the neutron star. As far as non-magnetic boundary layers are concerned, the spherical shape of the star should be taken into account as matter is expected to spread over the surface of the star \citep{IS99}. Although it is beyond the scope of the present work, the variability analysis of the funnel stream can, in principle, be done separately from the underlying disk using spherical coordinates to see the effects of the geometry and boundary conditions associated with the accretion stream on the appearance or disappearance of QPOs. In our picture, the density modulation of the accreting plasma along the funnel flow is directly fed by the global disk modes in the magnetic boundary layer \citep{EPA2008}. We therefore expect the frequency bands of the global modes (Section~\ref{qpomf}) that would yield the observed kHz QPOs to remain unaltered along the funnel stream.

\subsection{Concluding Remarks}\label{remark}

We proposed that the lack of correlation between kHz QPO frequency and X-ray flux in the long term intensity evolution of the source in neutron star LMXBs is a good indicator of the possible existence of quasi-steady states the source can occupy in the neighborhood of certain flux levels.

Each parallel track observed in the plane of $\nu_{\mathrm{kHz}}$ versus $F_{\mathrm{X}}$ corresponds to one of the possible quasi-steady states the magnetic boundary layer can reach near certain flux levels, at which the variation in $\dot{M}$ is sufficiently slow on short time scales (hours).

It is highly likely that rapid variations in $\dot{M}$ occur on long time scales (more than a day) and the source seeks a new quasi-steady state (a new track) for the boundary layer to generate kHz QPOs.

According to the proposed model, kHz QPO frequencies in a given source are mainly determined by the MHD properties of the innermost disk region such as the radial width of the boundary layer, the magnetospheric radius, and the magnetic diffusivity at the inner disk, which are all supposed to vary with $\dot{M}$.

We conclude from the comparison of the model predictions with the observed data of parallel tracks of individual sources such as 4U~1608--52, 4U~1636--53, and Aql~X-1 that the magnetic boundary layer evolves in accordance with the average steady behavior throughout the long-term flux evolution of the source. Short-term deviations from the average steady behavior are likely to be the result of variations in the magnetic diffusivity of the boundary layer, following reconnections between field lines.

\acknowledgments We thank the anonymous referee for valuable comments and suggestions that improved the manuscript. This work was supported by the Scientific and Technological Research Council of Turkey (T\"{U}B\.{I}TAK), under the project grant 114F100.

\end{document}